\documentstyle[11pt]{article}
\normalbaselines

\begin{document}
\bibliographystyle{unsrt}

\def\bea*{\begin{eqnarray*}}
\def\eea*{\end{eqnarray*}}
\def\ba{\begin{array}}
\def\ea{\end{array}}
\count1=1
\def\be{\ifnum \count1=0 $$ \else \begin{equation}\fi}
\def\ee{\ifnum\count1=0 $$ \else \end{equation}\fi}
\def\ele(#1){\ifnum\count1=0 \eqno({\bf #1}) $$ \else \label{#1}\end{equation}\fi}
\def\req(#1){\ifnum\count1=0 {\bf #1}\else \ref{#1}\fi}
\def\bea(#1){\ifnum \count1=0   $$ \begin{array}{#1}
\else \begin{equation} \begin{array}{#1} \fi}
\def\eea{\ifnum \count1=0 \end{array} $$
\else  \end{array}\end{equation}\fi}
\def\elea(#1){\ifnum \count1=0 \end{array}\label{#1}\eqno({\bf #1}) $$
\else\end{array}\label{#1}\end{equation}\fi}
\def\cit(#1){
\ifnum\count1=0 {\bf #1} \cite{#1} \else 
\cite{#1}\fi}
\def\bibit(#1){\ifnum\count1=0 \bibitem{#1} [#1    ] \else \bibitem{#1}\fi}
\def\ds{\displaystyle}
\def\hb{\hfill\break}
\def\comment#1{\hb {***** {\em #1} *****}\hb }

\newcommand{\TZ}{\hbox{\bf T}}
\newcommand{\MZ}{\hbox{\bf M}}
\newcommand{\ZZ}{\hbox{\bf Z}}
\newcommand{\NZ}{\hbox{\bf N}}
\newcommand{\RZ}{\hbox{\bf R}}
\newcommand{\CZ}{\,\hbox{\bf C}}
\newcommand{\PZ}{\hbox{\bf P}}
\newcommand{\QZ}{\hbox{\bf Q}}
\newcommand{\HZ}{\hbox{\bf H}}
\newcommand{\EZ}{\hbox{\bf E}}
\newcommand{\GZ}{\,\hbox{\bf G}}
\newcommand{\DZ}{\, \hbox{\bf D}}

\vbox{\vspace{38mm}}
\begin{center}
{\LARGE \bf Algebraic Geometry Approach to the Bethe 
Equation for  Hofstadter Type Models}\\[10 mm]
Shao-shiung Lin \\ {\it Department of Mathematics,\\ 
Taiwan University \\ Taipei, Taiwan \\ 
(e-mail: lin@math.ntu.edu.tw ) }\\[3 mm] Shi-shyr
Roan
\footnote{Supported in part by the NSC grant of Taiwan . }
\\{\it Institute of Mathematics \\ Academia Sinica \\ 
Taipei , Taiwan \\ (e-mail: maroan@ccvax.sinica.edu.tw)} 
\\[35mm]
\end{center}

\begin{abstract}
We study the diagonalization problem of certain
Hofstadter-type models through the algebraic Bethe
ansatz equation by the algebraic geometry method. When
the  spectral variables lie on a rational curve, we
obtain the complete and explicit solutions for  models
with the rational magnetic flux, and  discuss the Bethe
equation of their thermodynamic flux limit. The algebraic
geometry properties of the Bethe equation on  high genus
algebraic curves  are  investigated  in cooperation 
with physical considerations of the Hofstadter model.
\end{abstract}

$$
\begin{array}{ll}
PACS:& 02.10.{\rm Rn}, \ 03.65.{\rm Fd}, \ 5.30, \
75.10{\rm Jm}.
\\ Key \ words: &  {\rm  
Hofstadter \ model, \ \ Transfer \ matrix, \ \ Baxter \
vector,
\
\ Bethe
\ equation }
\end{array}
$$

\vfill
\eject

\section{Introduction}
The Hofstadter Hamiltonian is defined by 
\be
H_{Hof}  =  \mu (   \alpha U +
\alpha^{-1}
 U^{-1})  + \nu ( \beta V +
 \beta^{-1} V^{-1} )  ,
\ele(HofH)
where $U, V$ are unitary
operators satisfying the Weyl commutation relation for
an absolute value $1$ complex number $\omega$, $ U V =
\omega VU
$, and $\alpha,
\beta, \mu, \nu$ are parameters with
$|\alpha|=|\beta|=1$, $\mu, \nu \in \RZ$.
For  a primitive
$N$th root of unity $\omega \ (= e^{2 \pi
\sqrt{-1} \Phi})$,  i.e., the phase factor  
$\Phi = P/N$ with $P$ relatively prime to $N$,
one may assume the
$N$th power identity of $U, V$: $U^N=V^N=1$.
There are several physical interpretations for
the  Hofstadter model, especially in solid state
physics. A prominent one is to  consider it as a
tight-binding approximation for electrons bound
to atomic sites in a two-dimensional crystal and
in a strong external magnetic field. The history
of the  Hofstadter model can  trace back to the
work of Peierls
\cite{P} on Bloch electrons in metals with the
presence of a constant external magnetic field.
By the pioneering works of the 50s and 60s
\cite{A, C,  Har, L,  Z}, the role of magnetic
translations was found for the Hamiltonion
(\req(HofH)), where the Weyl pair of operators
, $\alpha U$ and $\beta V$, is a discrete version
of magnetic translations in the $x, y$-directions,
with the phase
$\Phi$ of their commutation factor  
$\omega $
representing  the magnetic flux (per plaquette),
and $\mu, \nu $ are the hopping amplitudes of the
system. Subsequently, a systematic study of
this 2D lattice model had begun. With the
discovery of quantum Hall effects, a large number of
interesting and important theoretical papers 
on Hamiltonions of the Hofstadter type
appeared in the eighties for the quantum
mechanical interpretation of the Hall
conductivity plateaus  (see, e.g.,
\cite{FN} and references therein). In 1976,
Hofstadter
\cite{H} found the butterfly figure of the
spectral band versus the magnetic flux, in which 
a beautiful fractal picture is exhibited. A
detailed study of the model and other
Hofstadter-type models began thereafter on the 
explanation on the fractal structure through
various mathematical approaches, such as the
semiclassical approximation, WKB-analysis on
difference equation, non-commutative geometry and
others \cite{ ATW, BS, BF,  CEY, HS, Wil}.  A
pedagogical account of this important aspect can
be found in a vast literature ( e.g., \cite{B,
Shu} and references therein). 

On the
other hand, motivated by the work of Wiegmann and
Zabrodin \cite{WZ}
 on the appearance of quantum $U_q(sl_2)$
symmetry in problems of magnetic translation, Faddeev
and Kashaev \cite{FK} pursued the diagonalization
problem on the following type of Hamiltonian by the
quantum transfer matrix  method 
developed by the Leningrad school in the early
eighties (cf, for instant,
\cite{TF}):
$$
H_{FK}= \mu ( \alpha U + \alpha^{-1} U^{-1} ) + \nu (
\beta V + \beta^{-1} V^{-1}) + \rho ( \gamma W +
\gamma^{-1} W^{-1} ) \ ,
$$
where
$U, V, W$ are unitary operators with the
Weyl commutation relations and the $N$th power
identity property, $UV=
\omega VU$, $ VW=
\omega WV$, $ WU= \omega UW$; $ U^N= V^N = W^N
=1$. With $\rho = 0$ in  $H_{FK}$,
one has the Hofstadter Hamiltonion (\req(HofH))
with the rational flux. Though the physical
content of the extended Hamiltonion $H_{FK}$ with
one more added operator
$W$ has not yet been clarified, this general
formulation will provide a nice setting in the
study of this kind of Hofstadter-type
Hamiltonions by the quantum inverse scattering
method. 
With the solution of Bethe ansatz equation under a
certain postulated degree condition (see (5.26) of
\cite{FK}), the energy expression of (\req(HofH))
obtained in
\cite{WZ} was reproduced in the $\rho=0$ limit
computation. Furthermore, a
general frame work of calculating spectra of the
Hamiltonian through the six-vertex model by the
algebraic Bethe ansatz method was presented in
\cite{FK}. In this approach, the Bethe ansatz equation is
formulated through the Baxter vector\footnote{It is also 
called as the "Baxter vacuum
state" in other literature. Here we follow the
terminology used in \cite{FK}. }
\cite{Bax, BazS}, visualized on a "spectral" curve 
associated to the corresponding Hofstadter-type
model.  In general, the spectral curve 
is a Riemann surface with a very high genus. Thus the
Bethe equation here can be viewed as a version of the
Baxter's T-Q relation \cite{Bax} on a
high genus spectral curve. The method relies on a special
local monodromy solution of the Yang-Baxter equation for
the six-vertex
$R$-matrix. This solution also appeared in
the study of chiral Potts model
\cite{BBP}. For a finite size
$L$, the trace of the $L$-monodromy matrix
gives rise to the transfer matrix acting on the
quantum space $\stackrel{L}{\otimes}\CZ^N$, where
$N$ is the order of the rational flux. The
transfer matrix is composed of a set of commuting
operators,  one of which is a Hofstadter-like
Hamiltonian. For the physical consideration, the 
size
$L$ is kept to be a fixed number. While motivated
by the Hofstadter butterfly spectral figure, 
the "thermodynamic" limit will be treated in a
manner of the  rational flux's order
$N$  tending to infinity. This process is a
different kind of limiting procedure
from many in quantum integrable systems, such as
the XYZ  spin chain or its degenerated forms. For
the study of models in this paper, we are mainly
concerned with the problem
of diagonalizing the transfer matrix of size
$L=3$ (see
\cite{FK}, or  Sects. 6, 7 of the
content). The Bethe ansatz
equations of the Hofstadter-like
Hamiltonians arising in such manner were clearly proposed
in \cite{FK}, and the principle could be equally applied
to a more general, possibly interesting, models.
However,  the explicit solutions and their
qualitative nature have  not yet been known, 
even in the situation when the spectral curve is a
rational one.  It is the object of
this paper to show that the algebraic geometry
method could provide an effective tool for a
thorough investigation on the mathematical
structure of the Bethe equation along the line of
the quantum transfer matrix scheme in \cite{FK}. 
In this work, we obtain the complete solutions of
the Bethe equation for models with the  rational
spectral curve for $L \leq 3$,  among which a
special kind of Hofstadter-type Hamiltonian
$H_{FK}$ is treated. We present a detailed and rigorous
mathematical derivation of these solutions of
Bethe ansatz equations, including the one in
\cite{FK}, and further expanding it to  all the
other sectors. Both the qualitative and
quantitative properties of the solutions are
studied thoroughly. With the understanding of
Bethe solutions for these Hofstadter-like models
of the rational flux, we apply the thermodynamic
process which  enables us to propose the Bethe
equation for a generic flux with the desirable
solutions from both the mathematical and physical
considerations. The analysis we make on the
solution of the model via the Bethe ansatz method
has clearly revealed the algebraic geometry
character of the Bethe equation. We adopt such an
interpretation in approaching the diagonalization
problem of certain quantum  integrable models.
Accordingly, the Bethe ansatz method for the Hofstadter
model (\req(HofH)) is 
formulated along this scheme, and the qualitative
nature of Bethe solutions is  obtained  
through the approach of algebraic curve theory.

This paper is organized as follows: 
In Sect. 2, we outline the concept of Baxter
vector, a key  ingredient for the study of the
Bethe (ansatz) equation throughout this work. In order
to gain the conceptual clarity, we shall present the
subject from the algebraic geometry aspect. In Sect. 3,
we  first recall some  results in
\cite{FK}  relevant to our later discussion,  fix 
notations on the transfer matrix and Bethe equation,
then derive  some general qualitative properties about
eigenvalues of the transform matrix. In the next four
sections, we shall consider the case when the spectral
data lie on a rational curve, and perform the
mathematical  derivation of the answer, along with the
discussion of their  physical applications. In Sect. 4, 
we briefly review the basic  procedure of limit
reduction to a rational spectral curve, 
originally appeared in
\cite{FK}, and present an explicit  form of Baxter
vector, which will be used in later discussions.
We present the complete solutions of  the Bethe
equations of all sectors for $L
\leq 3$ in Sect. 5.  In Sect. 6,  we explain the
"degeneracy" relations between the Bethe solutions and
the eigenspaces in the quantum space of the transfer
matrix. We also mention the relationship between
our Bethe solutions and some known results by
using the Bethe ansatz method in literature, and
indicate the difference of our approach of the Bethe
equation with the usual Bethe-ansatz-type technique
through certain non-physical solutions obtained by the
latter method.  In Sect. 7, we will address the
"thermodynamic" limit problem in the sense that  $N$
tends to
$\infty$, and  discuss the Bethe equation for those
Hofstadter-like Hamiltonions appeared previously
in the context of Sect. 5, but with  a
generic $q$.
In Sect. 8, we study the Hofstadter
Hamiltonian (\req(HofH)) by starting from the
format in
\cite{FK}. Then we go through  an algebraic
geometry analysis of the high genus spectral curve
in which solutions of the Bethe equation are
represented, and the  connection of the spectral
curves with elliptic curves are then clarified.
We obtain a primary understanding of Bethe
solutions through the spectral curve, and
their relation with the Heisenberg algebra.  We
end with the  concluding remarks in Sect. 9 with
a discussion of the directions of our further
inquiry. 

{\bf Convention .} In this paper, 
$\RZ, \CZ$ will denote 
the field of real, complex numbers respectively, and
$\ZZ$ the ring of integers.  For positive integers $N,
n$, we will denote
$\stackrel{n}{\otimes} \CZ^N$ the tensor
product of $n$-copies of the complex $N$-dimensional
vector space $\CZ^N$, and $\ZZ_N$ the quotient ring
$\ZZ/N\ZZ$.

\section{Algebraic Geometry Preliminary and Baxter
vector}  
In this paper, we shall denote $\omega$ a
primitive
$N$th root of unity, and $q =
\omega^{\frac{1}{2}}$ with   
$q^N = (-1)^{N+1}$, in particular for odd $N$,   
$q = \omega^{\frac{N+1}{2} }$.
Let $Z, X$ be two operators  
satisfying the Weyl commutation relation with the
$N$th power identity, 
$$
ZX= \omega XZ, \ \ \ \ 
Z^N = X^N = I \ .
$$  
The algebra generated by $Z,
X$ is called the Weyl algebra, in which the element $ZX$
will be denoted by $Y : =ZX$. The 
following  relations hold,
$$
XY = \omega^{-1} YX \ , \ \ \ YZ = \omega^{-1} ZY 
\ , \ \ \ Y^N= (-1)^{N-1} \ .
$$
The canonical irreducible 
representation of the Weyl algebra is given by the
following expressions on the $N$-dimensional space
$\CZ^N$,
\bea(lll)
Z : & v \mapsto Z(v) , &Z(v)_k = q^{2k} v_k \ ,
\\ X : &v \mapsto X(v) ,& X(v)_k = v_{k-1} \ , \
\\ Y : & v \mapsto Y(v) , &Y(v)_k = q^{2k}
v_{k-1} \ .
\elea(sRep)
Here a vector
$v $ of
$\CZ^N$ is represented by a sequence of coordinates, $v =
(v_k)_{k
\in \ZZ}$, with the $N$-periodic condition, $v_k =
v_{k+N}$. Hence one can consider the index $k$ as an
element of
$\ZZ_N$ in what follows if no confusion could arise. 
Denote $|k\rangle$
the standard basis of $\CZ^N$, $\langle k|$ the dual
basis of $\CZ^{N*}$ for $k \in \ZZ_N$ . The $k$-th
component of a vector
$v $ of
$\CZ^N$ is given by
$ v_k = \langle k | v \rangle$. 
In the Weyl algebra, we shall consider only the vector
subspace spanned by  $X, Y, Z$ and the
identity 
$I$, in which we denote the non-zero operator 
$$
\varphi_{\alpha, \beta, \gamma, \delta} : =
\alpha Y -
\beta X -
\gamma Z +
\delta I \ : \ \CZ^N \ \longrightarrow \CZ^N  \ .
$$
The kernel of the above operator, ${\rm
Ker}(\varphi_{\alpha, \beta, \gamma, \delta})$, depends
only on the ratio of the coefficients, i.e., the
element
$[\alpha,
\beta , \gamma, \delta] $ in the projective
3-space $\PZ^3$. The non-triviality of 
${\rm Ker}(\varphi_{\alpha, \beta, \gamma, \delta})$
defines a hypersurface of $\PZ^3$, 
$$
{\cal F} := \{ [\alpha, \beta, \gamma, \delta] \in
\PZ^3 
\ | \ {\rm Ker}(\varphi_{\alpha, \beta, \gamma,
\delta}) \neq 0 \} \ .
$$
In fact, one has the defining equation of ${\cal
F}$ as follows:
\newtheorem{lemma}{Lemma}
\begin{lemma}\label{lem:ker}
The surface ${\cal F}$ is defined by the
equation,
$$
{\cal F} : \   \alpha^N + \delta^N = \beta^N  +
\gamma^N 
\ , \ 
\  \ [\alpha, \beta, \gamma, \delta ] \in \PZ^3 \ . 
$$ 
Furthermore for $[\alpha,
\beta ,
\gamma,
\delta] \in {\cal F}$,   
${\rm Ker}(\varphi_{ \alpha , \beta , \gamma ,
\delta })$ is an one-dimensional subspace of
$\CZ^N$ generated by a vector $v = (v_m)$ with the
ratios, 
$ v_m : v_{m-1} = ( \alpha \omega^m  - \beta ): (
\gamma \omega^m - \delta )$.  
\end{lemma}
{\bf Proof. } For $v \in \CZ^N$, it is obvious that the
criterion of $v$ in  
${\rm Ker}(\varphi_{ \alpha , \beta , \gamma ,
\delta })$  is described by the above
ratio relations of
$v_m$s. 
For a non-zero vector  $v$ of $\CZ^N$, the
periodic relation 
$v_m = v_{m+N}$ for a non-zero component $v_m$
gives rise to the equation of ${\cal F}$. 
$\Box$ \par \vspace{0.2in} \noindent
Let $v$ be a basis of ${\rm Ker}
(\varphi_{ \alpha , \beta , \gamma , \delta })$ for 
$[\alpha , \beta , \gamma , \delta ] \in {\cal
F}$. Note that there are $N^2$ (projective) lines in  
${\cal F}$,  defined by 
$\alpha^N-\beta^N=0$, equivalently,  
$\gamma^N -
\delta^N=0$, which are labelled by $\PZ^1_{j, k}
:= \{   \alpha \omega^j -\beta= 
\gamma- \delta \omega^k =0
  \} $ for $j, k
\in \ZZ_N$.
Outside these lines, the component of $v$ are
all non-zero. For elements in $\PZ^1_{j,k}$, one can
set $v = |k\rangle$ at $[0, 0, \omega^k , 1]$, and 
$v = |j-1\rangle$ at $[1, \omega^j, 0, 0]$; while for the
rest of elements, the indices in
$\ZZ_N$ with  the non-zero components of $v$ 
form a chain from $k$ increasing to $j-1$.

For the later purpose, we introduce a family of
non-homogenous  representations  of the
surface ${\cal F}$, depending on the parameter $
h =[a, b, c, d] \in \PZ^3$,
$$
\alpha = \xi'a \  , \ \ 
\beta=  xb \ , \ \
\gamma=  -  \xi' \xi x c \
, \ \ 
 \ \delta= - \xi d \ ,
$$ 
where $(x, \xi, \xi') \in \CZ^3$ satisfies the
equation,
\begin{eqnarray*}
{\cal S}_h \ : \ \ \  \xi'^Na^N - x^Nb^N
 = (-\xi)^N( x^N \xi'^N c^N - d^N )  \ .
\end{eqnarray*}
For a fixed $h \in \PZ^3$, the 
operator
$\varphi_{\alpha ,
\beta , \gamma ,
\delta }$ corresponding to $(x, \xi, \xi')$ becomes
\bea(l)
F(x,  \xi, \xi' ) ( = F(x,  \xi, \xi' ; h)) :=
\xi'aY - xbX +  \xi'
\xi xc Z -\xi d I  \ ,
\elea(def:F)
and we have
\begin{eqnarray*}
F(x, \xi-1, \xi') = F(x,  \xi, \xi' )- 
\xi' x cZ+d I \ , \ \  &  
F(x,  \xi, \xi'-1 ) = F(x, 
\xi, \xi' ) -  \xi x cZ - aY \ . 
\end{eqnarray*}
For an element $p=(x, \xi, \xi')$ of ${\cal S}_h$, we
shall denote $|p\rangle$ the basis of $ {\rm
Ker}(\varphi_{
\xi'a, xb , -  \xi' \xi x c , - \xi d })$ with $
\langle 0| p\rangle = 1 $, 
equivalently, $|p\rangle$ is the vector of $\CZ^N$ 
defined by  
$$
\langle 0| p\rangle = 1 \ , \ \ \ \
\frac{\langle m|p\rangle}{\langle m-1|p\rangle} = 
\frac{\xi'a
\omega^m  - xb }{
- \xi (  \xi' x c \omega^m - d) }  \ .
$$
We shall call  $|p\rangle$  the Baxter vector 
associated to $p \in {\cal S}_h$ \cite{Bax, BazS, FK}.
Then
\bea(cll)
F(x, \xi, \xi' )|p\rangle &= &\vec{0} \ , \\
F(x, \xi -1, \xi' ; h )|p\rangle &= &  |\tau_- p\rangle 
\Delta_-(p) , \\
F(x, \xi, \xi'-1,; h)|p\rangle &= & - |\tau_+
p\rangle\Delta_+(p) \ ,
\elea(Baxtv)
where $\Delta_\pm$ are the following (rational) 
functions of ${\cal S}_h$,
$$
\Delta_-(x, \xi, \xi')  =  d-x \xi' c
\ , \ \ \Delta_+(x, \xi, \xi') = 
 \frac{\xi (ad-x^2bc)}{\xi'a -xb}  ,
$$
and $\tau_{\pm}$ are the automorphisms of 
${\cal S}_h$  defined by $
\tau_\pm  (x, \xi, \xi') 
= (q^{\pm 1} x, 
q^{-1} \xi, q^{-1} \xi') $.

\section{Transfer Matrix and the Bethe Equation}
As in \cite{FK}, it is known that the following 
$L$-operator\footnote{ The
conventions in this paper are different from the ones
used in  \cite{FK} where $X, Y$ correspond to our
$Z, X$ here.} with the operator-valued entries
acting on the quantum space
$\CZ^N$,    
$$
L_h (x) = \left( \begin{array}{cc}
       aY  & xbX  \\
        xcZ &d    
\end{array} \right) \ , \ \ x \in \CZ \ ,
$$
possesses the intertwining property of the Yang-Baxter
relation, 
\begin{eqnarray}
R(x/x') (L_h (x) \bigotimes_{aux}1) ( 1
\bigotimes_{aux} L_h (x')) = (1
\bigotimes_{aux} L_h (x'))(L_h (x)
\bigotimes_{aux} 1) R(x/x') \ , \label{eq:RLL}
\end{eqnarray}
where the script letter $"aux"$ will always indicate an 
operation taking on the auxiliary space $\CZ^2$, 
$R(x)$ is the matrix of a 2-tensor of the auxiliary 
space  with the following  numerical expression,
$$
R(x) = \left( \begin{array}{cccc}
        x\omega-x^{-1}  & 0 & 0 & 0 \\
        0 &\omega(x-x^{-1}) & \omega-1 &  0 \\ 
        0 & \omega-1  &x-x^{-1} & 0 \\
     0 & 0 &0 & x\omega-x^{-1}     
\end{array} \right) \ .
$$
By performing the matrix product on 
auxiliary  spaces and the tensor product of quantum 
spaces,
one has the $L$-operator associated to an element 
$\vec{h}= (h_0,
\ldots, h_{L-1}) \in  (\PZ^3)^L $,
\begin{eqnarray}
L_{\vec{h}}(x) =
\bigotimes_{j=0}^{L-1}L_{h_j}(x) := L_{h_0}(x)
\otimes L_{h_1}(x) 
\otimes \ldots \otimes L_{ h_{L-1}}(x)  \ , \label{tLij}
\end{eqnarray}
which again satisfies the relation 
(\ref{eq:RLL}). The entries of $L_{\vec{h}}(x)$ are
operators of the quantum space 
$\stackrel{L}{\otimes} \CZ^N$, and its trace defines the
transfer matrix, 
$$
T_{\vec{h}} (x) = {\rm tr}_{aux} ( L_{\vec{h}}(x))
\ .
$$
Then the commutation relation holds,
$$
[ T_{\vec{h}} (x)  , T_{\vec{h}} (x')  ] = 0 \ , \ \ \
\ x , x' \in \CZ \ .
$$
The transfer matrix $T_{\vec{h}} (x)$ can also be
computed by changing
$L_{h_j}$ to
$\widetilde{L}_{h_j}$ via a gauge transformation in the
following manner,
\bea(l)
\widetilde{L}_{h_j}(x) = A_j L_{h_j}(x) A_{j+1}^{-1}
\ ,\ \ \ A_{L}: = A_0 \ , \ \ \ 
0 \leq j \leq L-1 \  .
\elea(gauge)
Set 
$$
A_j = \left( \begin{array}{lc}
1 &  \xi_j-1\\
1 &  \xi_j 
\end{array}\right) \ ,
$$
and denote the corresponding $\widetilde{L}_{h_j}(x)$ by
$
\widetilde{L}_{h_j} (x,
\xi_j, \xi_{j+1})$. With $F_h(x, \xi, \xi')$ of
(\req(def:F)), we have 
$$
\widetilde{L}_{h_j} (x, \xi_j, \xi_{j+1})  = \left(
\begin{array}{ll} F_{h_j}(x,  \xi_j -1, \xi_{j+1}) &
-F_{h_j} (x, 
\xi_j -1, \xi_{j+1} -1) \\ F_{h_j}(x,  \xi_j ,
\xi_{j+1}) & - F_{h_j}(x, 
\xi_j ,
\xi_{j+1}-1 )
\end{array}\right) \ ,
$$
and 
$$
T_{\vec{h}} (x) = {\rm tr}_{aux}
(\widetilde{L}_{\vec{h}}(x,
\vec{\xi} ) ) \ , \ \ \vec{\xi} := 
(\xi_0,  \ldots ,
\xi_{L-1} )
$$
where 
\begin{eqnarray*}
\widetilde{L}_{\vec{h}}(x,
\vec{\xi} ) : = 
\bigotimes_{j=0}^{L-1}
\widetilde{L}_{h_j}(x,
\xi_j, \xi_{j+1})   = 
\left( \begin{array}{cc}
   \widetilde{L}_{ \vec{h}; 1,1}(x, \vec{\xi}) &
\widetilde{L}_{
\vec{h};1,2}(x, \vec{\xi})  \\
     \widetilde{L}_{ \vec{h}; 2,1}(x, \vec{\xi}) &
\widetilde{L}_{ \vec{h} ; 2,2}(x, \vec{\xi}) 
\end{array} \right) , \ \ \ \ \xi_L := \xi_0 \ .
\end{eqnarray*}
The existence of Baxter vectors $|p_j\rangle$, $ 
p_j := (x, \xi_j,
\xi_{j+1}) \in {\cal S}_{h_j}$ for all $j$ with
the condition
$\xi_L = \xi_0$, imposes the constraint of
elements $(p_0, \ldots, p_{L-1})$ on the product
of surfaces, 
$\prod_{j=0}^{L-1}{\cal S}_{h_j}$, which form a
curve 
${\cal C}_{\vec{h}}$ in $\prod_{j=0}^{L-1}{\cal S}_{h_j}$
with the coordinates
$(x, \xi_0,
\ldots, \xi_{L-1})$ satisfying the
relations,
\begin{eqnarray}
{\cal C}_{\vec{h}} : \ \ \xi_j^N  =(-1)^N
\frac{\xi_{j+1}^Na_j^N - x^Nb_j^N}{\xi_{j+1}^N x^N
c_j^N - d_j^N } \ \ , \ \ \ \  j =0,
\ldots, L-1 . \label{eq:Cvh}
\end{eqnarray} 
For $p= (p_0, \ldots, p_{L-1}) \in {\cal
C}_{\vec{h}}$, the Baxter vector 
$|p\rangle$ is now defined by 
$$
|p\rangle \ : = |p_0\rangle \otimes \ldots \otimes
|p_{L-1}\rangle 
\in \ \ \stackrel{L}{\otimes} \CZ^N \ .
$$
By the definition of
$\widetilde{L}_{\vec{h}; j,k}$, the Baxter vector of $ 
{\cal C}_{\vec{h}}$  shares
the following relations to
entries of $\widetilde{L}_{\vec{h}}$ similar to  those
for $\widetilde{L}_{\vec{h}}(x, \vec{\xi})$ in
(\req(Baxtv)), 
\begin{eqnarray*}
\widetilde{L}_{\vec{h}; 1,1}(x, \vec{\xi})|p\rangle = 
|\tau_- p\rangle 
\Delta_-(p) , &
\widetilde{L}_{\vec{h}; 2,2}(x, \vec{\xi})|p\rangle = 
|\tau_+ p\rangle\Delta_+(p) \ , &
\widetilde{L}_{\vec{h}; 2,1}(x, \vec{\xi})|p\rangle = 0
\ , 
\end{eqnarray*}
where $\Delta_\pm, \tau_\pm$ are the functions
and  automorphisms of 
${\cal
C}_{\vec{h}}$ defined by
\begin{eqnarray}
&\Delta_-(x, \xi_0, \ldots, \xi_{L-1})  &=
\prod_{j=0}^{L-1}( d_j-x
\xi_{j+1} c_j ) \ , \nonumber \\
&\Delta_+(x, \xi_0, \ldots, \xi_{L-1}) &= 
\prod_{j=0}^{L-1} \frac{\xi_j
(a_jd_j-x^2b_jc_j)}{\xi_{j+1}a_j -xb_j} \  , 
\label{DelTau} \\
&\tau_\pm :  (x, \xi_0, \ldots, \xi_{L-1}) 
&\mapsto (q^{\pm 1} x, 
q^{-1} \xi_0, \ldots, q^{-1} \xi_{L-1}) \ 
\nonumber .
\end{eqnarray}
Then follows the important relation of 
the transfer matrix on the Baxter vector of the curve 
${\cal C}_{\vec{h}}$, 
\bea(l)
T_{\vec{h}}(x) |p\rangle = |\tau_- p\rangle 
\Delta_-(p)  + |\tau_+ p\rangle\Delta_+(p) \ , \ \ 
{\rm for } \ \ p \in {\cal C}_{\vec{h}} \ .
\elea(T|p)
As $T_{\vec{h}}(x)$ are commuting operators for
$x \in \CZ$, 
a common eigenvector $\langle \varphi|$ is a constant 
vector of $\stackrel{L}{\otimes} \CZ^N$ 
with an eigenvalue
$\Lambda(x) \in \CZ[x]$. Define the function
$Q(p)= \langle \varphi|p\rangle$ of ${\cal
C}_{\vec{h}}$, then it 
 satisfies the
following Bethe equation, 
\bea(l)
\Lambda(x) Q(p)  = Q(\tau_-(p)) \Delta_-(p) 
+  Q(\tau_+(p)) \Delta_+(p) \ , \ \ {\rm for} \ p 
\in {\cal C}_{\vec{h}} \ .
\elea(Bethe)
We are going to make a detailed study of the above Bethe
equation for the rest of this paper. Before that,
 we first 
derive certain functional properties 
of the eigenvalue $\Lambda (x)$.
\begin{lemma}\label{lem:Lij}
With the entries $L_{\vec{h};i,j}(x)$ 
of $L_{\vec{h}}(x)$ in {\rm (\ref{tLij})}, the following
properties hold. 

(I) Under the
interchange of operators,
$a_jY \leftrightarrow d_j$ ,  $b_jX 
\leftrightarrow c_jZ $ for all $j$, 
we have the symmetries among  $L_{\vec{h};i,j}(x)$s, \  
$L_{\vec{h};1,1}(x) \leftrightarrow L_{\vec{h}; 2,2}(x)$
, $L_{\vec{h};1,2}(x) \leftrightarrow  L_{\vec{h};
2,1}(x) $.

(II) The polynomial 
$L_{\vec{h};i,j}( x) $ is an even or odd
function with the parity $(-1)^{i+j}$, and its degree
is equal to $  2[\frac{L+1-\delta_{i,j}}{2}]-1+
\delta_{i,j}$.
\end{lemma}
{\bf Proof. } We apply the gauge transformation
(\req(gauge)) with $A_j=A_{j+1}$ for all $j$. When
$$
A_j = \left( \begin{array}{cc}
      0 & 1 \\
       1 &0 
\end{array} \right) \ , 
$$
one obtains the interchange of entries of
$L_{h_j}(x)$, hence follows $(I)$. For 
$$
A_j= \left( \begin{array}{cc}
      1 & 0 \\
       0 &-1 
\end{array} \right) \ , 
$$
the corresponding $\widetilde{L}_{h_j}(x)$ is equal to $
L_{h_j}(-x)$, which  implies the parity of
$L_{\vec{h};i,j}( x) $. The  determination of the
degree  of $L_{\vec{h};i,j}( x) $  can be obtained by
certain  suitable choices of the 
values of $h_j$s.  
$\Box$ \par \vspace{0.2in} \noindent
From the definition of $T_{\vec{h}}(x), 
\Lambda(x)$, one can easily obtain the following result.
\newtheorem{proposition}{Proposition}
\begin{proposition}\label{prop:Lambda}
The transform matrix 
$T_{\vec{h}}(x)$ is an operator-coefficient even
polynomial  of $x$  with degree $2[\frac{L}{2}]$, which
is invariant under the substitutions in Lemma
$\ref{lem:Lij} \ (I)$.  The constant term of
$T_{\vec{h}}(x)$ is given by 
\bea(l)
T_0 := T_{\vec{h}}(0) =  \prod_{j=0}^{L-1} a_j 
\stackrel{L}{\bigotimes} Y + 
\prod_{j=0}^{L-1} d_j \ .
\elea(T0)
Subsequently, the polynomial $\Lambda (x)$ in {\rm
(\req(Bethe))} is an  even function of
degree
$\leq 2[\frac{L}{2}]$ with $\Lambda (0) = 
q^l \prod_{j=0}^{L-1} a_j + \prod_{j=0}^{L-1} d_j  $ 
for some $l$.
\end{proposition}
By the above proposition, we have  $T_{\vec{h}}(x) =
\sum_{j=0}^{[\frac{L}{2}]} T_{2j} x^{2j}$ with  
$T_0$ given by (\req(T0)). With a further study
of the expressions of
$T_{2j}$s, one can show that they form a
commuting family  of operators, whose proof we won't 
present here. Instead, as an illustration of this fact
and also  for the later use of this paper, we list
below the explicit form of $T_2$ for $L=2, 3$ in  
$T_{\vec{h}}(x)= T_0 + x^2 T_2 $, where the
commutation relation of $T_2$ and 
$T_0$ is easily verified, 
\bea(lll)
L=2 , & T_2&= b_0c_1 X\otimes Z + c_0b_1 Z \otimes X \ ;
\\
 L=3 , & T_2 & = b_0c_1a_2 X\otimes Z \otimes Y +
a_0b_1c_2 Y \otimes  X \otimes Z+ 
 c_0a_1b_2 Z \otimes Y
\otimes X \\
 &  & + \ c_0b_1d_2 Z\otimes X \otimes I +
d_0c_1b_2 I \otimes  Z \otimes X+ 
 b_0d_1c_2 X \otimes I
\otimes Z \ .
\elea(T02)
The above $T_2$ for $L=3$ can be written as the
 Hofstadter-type Hamiltonian $H_{FK}$ of
Faddeev and Kashaev described in Sect. 1 ( for
the exact identification, see
\cite{FK}) \footnote{The operators $X, Y, Z, S, T, U$ in
\cite{FK} are corresponding to $Z, X, Y, U, V, W$
respectively here in this paper.}.
In the equation (\req(Bethe)), $Q(p)$ is a
 rational function  of ${\cal C}_{\vec{h}}$
with poles. As the functions of 
${\cal C}_{\vec{h}}$, 
$\xi_{j+1}^Nx^Nc_j^N-d_j^N$ and
$(\xi_{j+1}^Na_j^N-x^Nb_j^N)(-\xi_j)^{-N}$, are the
same, by the description of the 
Baxter vector, the poles of
$Q(p)$ are  contained in the divisor of 
${\cal C}_{\vec{h}}$ defined by 
$$
\prod_{j=0}^{L-1}
( \xi^N_jx^Nc_j^N - d_j^N) = 0 \ .
$$
Hence the understanding  of the Bethe solutions of
(\req(Bethe)) relies heavily on the function theory of
${\cal C}_{\vec{h}}$, the algebraic geometry of the
curve 
will play a key role on the complexity of the problem.
We  shall specify the spectral
curves in the later discussion.  For our purpose, in the
rest of this paper we shall only consider the situation
when $N$ is an odd integer, and denote the integer
$[\frac{N}{2}]$ by $M$,  
$$
N = 2M+1 \ .
$$

\section{The Rational Degenerated Bethe Equation}
For the next four sections, the spectral curve 
${\cal C}_{\vec{h}}$ will always be the rational curve
under the following assumption of degeneration,
\be
a_j = q^{-1} d_j \ , \ \ b_j = q^{-1} c_j \  \ \ \ \ \
{\rm for}
\  j=0, \ldots, L-1 \ .
\ele(rataspm)
To make our presentation self-contained, 
 we shall briefly review in this section the
general procedure of reducing the Bethe equation
on ${\cal C}_{\vec{h}}$ to a polynomial equation,
originally appeared in \cite{FK}, and present an
explicit form of the Baxter vector which will be
convenient for the later use. 
By replacing $c_j,
d_j$  by $\frac{c_j}{d_j}, 1$, we 
may assume $ d_j=1$ for all $j$.  
For the convenience of mathematical discussion, also
suitable for physical  applications,  we shall assume
that the parameters $c_j$s are all generic. 
The solutions for $\xi_j$s in (\ref{eq:Cvh}) are
given by 
$$
\xi_0^N=  \cdots = \xi_{L-1}^N = \pm 1 \ ,
$$
which possess the structure of a finite abelian group.
Therefore ${\cal C}_{\vec{h}}$ is the 
union of disjoint copies of the $x$-(complex) line
indexed by this finite group. Instead
working on the curve ${\cal C}_{\vec{h}}$, the
following
$\tau_\pm$-invariant subset of ${\cal C}_{\vec{h}}$ will
be sufficient for our discussion of the Bethe equation,
$$
{\cal C} := \{ (x, \xi_0, \ldots, \xi_{L-1}) \ | \ 
 \xi_0 =
\cdots =
\xi_{L-1} =  q^l \ , \ l \in \ZZ_N \} \
.
$$ 
The curve ${\cal C}$ will be identified with $\PZ^1
\times \ZZ_N$:
$$
{\cal C} = \PZ^1 \times \ZZ_N \ , \ \ \ \ 
(x,  q^l, \ldots,  q^l)
\longleftrightarrow  (x, l) \ .
$$
The Baxter vectors are now labelled by  
$|x, l \rangle = \otimes_{j=0}^3 |x, l \rangle_j
$, where $|x, l \rangle_j \in \CZ^N$ is defined by the
relations,
\begin{eqnarray*}
\langle 0|x, l \rangle_j = 1 \ , \ \ \ \
\frac{\langle k|x, l \rangle_j}{\langle k-1|x, l
\rangle_j} = 
\frac{ q^{2k-1} ( 1-xc_jq^{-l-2k} ) }{ 
( 1- xc_j q^{l+2k} ) }  \ , \ \ \ \  k \in \ZZ_N \ . 
\end{eqnarray*}
We shall use the bold letter ${\bf k}$ to denote a
multi-index vector
${\bf k}= (k_0, \ldots, k_{L-1})$ with
$k_j \in \ZZ$; the square-length of ${\bf k}$ is
defined by 
$|{\bf k}|^2:= \sum_{j=0}^{L-1} k_j^2$. The 
component-expression of the vector $|x, l \rangle$ is
given by
\begin{eqnarray}
\langle {\bf k}|x, l \rangle  = q^{|{\bf k}|^2}
\prod_{j=0}^{L-1} \prod_{i=1}^{k_j}
\frac{   1- xc_jq^{-l-2i} }{ 1-  xc_j q^{l+2i}  }
 \ , \ \ \ \ k_j > 0 \  \ .  \label{eq:Bvec}
\end{eqnarray}
The equation (\req(T|p)) takes the
form,
\bea(l)
T (x) |x, l \rangle = |q^{-1}x, l-1\rangle 
\Delta_-(x, l)  + |qx, l-1\rangle\Delta_+(x, l) \  , 
\elea(T|xm)
where $\Delta_\pm$ are given by 
$$
\Delta_-(x, l)  =
\prod_{j=0}^{L-1}( 1-x
c_j q^l ) \ , \ \ \ 
\Delta_+(x, l) = 
\prod_{j=0}^{L-1} \frac{ 1-x^2 c^2_j}{ 1 - 
x c_j q^{-l}} \   .
$$
As in \cite{FK}, we introduce the following two
functions on the curve ${\cal
C}$ ,
\begin{eqnarray}
f^e(x, 2n) =  \prod_{j=0}^{L-1}
\prod_{k=0}^{n} 
\frac{1-xc_jq^{-2(n-k)}}{1-xc_jq^{2(n-k)}} \ , \
\ \ \ 
f^o(x, 2n+1) =  \prod_{j=0}^{L-1}
\prod_{k=0}^{n} 
\frac{1-xc_jq^{-1-2(n-k)}}{1-xc_jq^{1+2(n-k)}} \ ,
 \label{eq:f}
\end{eqnarray}
by which we define the vectors,
\begin{eqnarray}
|x\rangle_m^e = \sum_{n=0}^{N-1}  |x, 2n\rangle  
f^e (x, 2n)
\omega^{mn} 
 \ , \ \ \
|x\rangle_m^o = \sum_{n=0}^{N-1} 
 |x, 2n+1\rangle f^o (x, 2n+1)  \omega^{mn}  \ .
\label{eq:xeo}
\end{eqnarray}
By computations, the following relations
hold for $n \in \ZZ_N$, 
$$
\frac{f^e (x, 2n) }{f^o (q^{\pm
1}x, 2n-1) }\Delta_{\pm}(x, 2n)  =
\Delta_{\pm}(x, 0) \ , \ \ \
\frac{f^o(x, 2n-1)}{f^e(xq^{\pm 1},
2n-2)}
\Delta_\pm(x, 2n-1) =
\Delta_\pm(x, - 1) \ .
$$
Then (\req(T|xm)) becomes the system of 
equations,
$$
T(x)|x\rangle\rangle_m
=|q^{-1}x\rangle\rangle_m D^-_m(x)  +
|qx\rangle\rangle_m D^+_m(x) \ , \ \ \ m \in \ZZ_N \ ,
$$
where  
$$
\begin{array}{lll}
|x\rangle\rangle_m=( |x\rangle_m^e , |x\rangle_m^o ) \
,  & D^{\pm}_m (x) = 
\left( \begin{array}{cc}
        0  & \Delta_{\pm}(x, -1)  \\
        \omega^m \Delta_\pm (x, 0) &0     
\end{array} \right) \ .
\end{array}
$$
In what follows, it is convenient to use the
notation of shifted factorial,
$$
(a ; \alpha )_0 = 1, \ \ \ \
(a ; \alpha )_n = 
(1-a)(1-a \alpha ) \cdots (1-a
\alpha^{n-1}) , \ \ \ \ n \in \ZZ_{>0} \ .
$$
By
(\ref{eq:Bvec}) (\ref{eq:f}), we
have 
\begin{eqnarray*}
f^e(x, 2n) \langle  {\bf k} |x,
2n \rangle     &=& q^{|{\bf
k}|^2}
\prod_{j=0}^{L-1} 
\frac{(xc_j ; \  \omega^{-1})_{k_j+n+1} }{(xc_j ;
\ \omega  )_{k_j+n+1} }  \ , \\ 
f^o(x, 2n+1) \langle {\bf k}|x, 2n+1 \rangle& = &
q^{|{\bf k}|^2}
\prod_{j=0}^{L-1} 
\frac{(xc_jq^{-1} ; \ \omega ^{-1})_{k_j+n+1}
}{(xc_jq ; \ \omega )_{k_j+n+1} } 
\ \ .  
\end{eqnarray*}
Note that each ratio-term  in the above right
hand side is defined when the lower index $k_j+n+1$ is
positive, however its value depends only the class
modular $N$. So we shall use the same notation for an
arbitrary integer index $n$ by defining the value
equal to that of any positive representative of the class
involved in 
$\ZZ_N$, and will keep this convention in what
follows. By (\ref{eq:xeo}), we have
\begin{eqnarray}
&\langle {\bf k}|x\rangle_m^e & = q^{|{\bf
k}|^2} \sum_{n\in \ZZ_N}
\omega^{mn} 
\prod_{j=0}^{L-1} 
\frac{(xc_j ; \ \omega^{-1})_{k_j+n+1} }{(xc_j ;
\ \omega )_{k_j+n+1} } \ , \nonumber \\ 
&\langle {\bf k}|x\rangle_m^o & =  q^{|{\bf k}|^2} 
\sum_{n \in \ZZ_N}  \omega^{mn} 
\prod_{j=0}^{L-1} 
\frac{(xc_jq^{-1} ; \ \omega^{-1})_{k_j+n+1}
}{(xc_jq ; \ \omega )_{k_j+n+1} } \ . 
 \label{formula}  
\end{eqnarray}
We proceed the  diagonalizing  procedure on  the matrix
$D^{\pm}_m (x)$  by a gauge transformation using  
an invertible $2 \times 2$
matrix $U_m(x)$, 
\begin{eqnarray*}
|x\rangle\rangle_m \mapsto |x\rangle_m := 
|x\rangle\rangle_m U_m(x)  \ , \ \ \  & 
D_m^{\pm}(x) \mapsto 
 U_m(q^{\pm
1}x)^{-1} D^{\pm}_m (x) U_m(x) \ .
\end{eqnarray*}
We shall choose the  matrix
$U_m(x)$ of the form \footnote{ The form of the
gauge transformation matrix here is 
slightly different from that of (5.15) in 
\cite{FK} by rearrangement of the entries. As we
are  not able to produce the  required formulation
through the expression in \cite{FK}, we wonder that there
might be a misprint there.}
$$
U_m(x) = \left( \begin{array}{cc}
         q^{-m}u(qx) &u(qx)   \\
        u(x)   & -q^mu(x)   
\end{array} \right) \ .
$$
Then the diagonalizable criterion of $D_m^\pm(x)
$ for all $m$  is equivalent to the following
equation of $u(x)$,
\begin{eqnarray}
\frac{u( \omega x)}{u(x)} = \prod_{j=0}^{L-1}
\frac{1 -c_jx}{1-c_jx q} \ .
\label{gaugeE}
\end{eqnarray}
Note that the right hand side of the above form is
equal to 
$\frac{\Delta_+(x, -1)}{\Delta_+(x, 0)}$, which is the
same as 
$ 
\frac{\Delta_-(qx, -1)}{\Delta_-(qx, 0)}$.
The resulting expression of  
$D_m^{\pm}(x)$ becomes
$$
D_m^-(x) = q^m \Delta_-(x, -1) \left(
\begin{array}{cc}
         1 &0 \\
        0   & -1   
\end{array} \right) \ , \ \ 
D_m^+(x) = q^m \Delta_+(x, 0) \left(
\begin{array}{cc}
         1 &0 \\
        0   & -1   
\end{array} \right) \ .
$$
With the notations
\begin{eqnarray*}
|x \rangle_m = (| x\rangle_m^+ , |
x\rangle_m^- ) \ , \ \ \ & (Q_m^+(x), Q_m^-(x) ) =
\langle
\varphi | x\rangle_m \ , 
\end{eqnarray*}
one has 
\begin{eqnarray*}
|x \rangle_m^+ =  |x\rangle_m^e q^{-m} u(qx) +  
|x \rangle_m^o u(x) \ , & |x\rangle_m^-  = 
|x \rangle_m^e u(qx) - |x\rangle_m^o q^mu(x)  \ .
\end{eqnarray*}
The Bethe equation (\req(Bethe)) now takes 
the form
\begin{eqnarray}
\pm q^{-m} \Lambda (x) Q_m^{\pm} (x) = 
\prod_{j=0}^{L-1}(1-xc_jq^{-1})
Q_m^{\pm} ( xq^{-1}) + \prod_{j=0}^{L-1} 
(1+x c_j) Q_m^{\pm} (xq) \ .
\label{QmEq}
\end{eqnarray}

\section{ Solutions of the Rational Bethe
Equation }
In this section, we advance the discussion 
of the last section to obtain the explicit solutions
of  the Bethe equation (\ref{QmEq}) for
$L \leq 3$, among which a special case will be
shown in the next section to coincide with the
one in \cite{FK, WZ}.
\begin{lemma}\label{lem:usol}
The general solutions of the rational function
$u(x)$ for {\rm (\ref{gaugeE})} are  given by 
$$
u(x) =  R(x^N) \prod_{j=0}^{L-1}(c_j x ; \
\omega)_{M+1}^{-1}  \ ,
$$
where $R(x^N)$ is a rational function of $x^N$.
\end{lemma}
{\bf Proof. } Note that the ratio of any two solutions of
(\ref{gaugeE}) is a rational function 
$r(x)$ with the relation
$r(\omega x)=r(x)$, which is equivalent to  
$r(x) = R(x^N)$ for a rational function
$R(x^N)$ of $x^N$. Therefore it suffices to show
that  $
 \prod_{j=0}^{L-1}(c_j x ; \
\omega)_{M+1}^{-1} $ is a solution of 
(\ref{gaugeE}), which is easily seen by 
$q=\omega^{M+1}$.
$\Box$ \par \vspace{0.2in} \noindent
By the expressions of $\langle {\bf k}|x \rangle_m^e ,
\langle {\bf k}|x\rangle_m^o $ in (\ref{formula}), in
order to have the  polynomial functions of 
$Q_m^{\pm}(x)$ in (\ref{QmEq}), we
choose the following gauge function $u(x)$ by
setting $R(x^N)= \prod_{j=0}^{L-1}  (1 - x^Nc_j^N)^2 $ 
in Lemma \ref{lem:usol},
\begin{eqnarray}
u(x)= \prod_{j=0}^{L-1} (1-x^Nc_j^N)(xc_jq ;
q^2)_M \ . \label{udef}
\end{eqnarray}
The  polynomial
$Q_m^{\pm}(x)$ has the degree  at most equal to that of $
u(x)$, which is $(3M+1)L$.  By Proposition
\ref{prop:Lambda}, one requires the polynomial
solutions 
$Q^\pm_m (x), \Lambda(x)$ of the Bethe
equation (\ref{QmEq}) with the constraints, 
$$
\begin{array}{lll}
{\rm deg.}Q^\pm_m (x)
\leq  (3 M +1)L &&\\
 {\rm deg.}
\Lambda (x) \leq 2[\frac{L}{2}], &
\Lambda (x)= \Lambda(-x), & \Lambda (0)= 
q^l+1  {\rm \ for \ some \ } l.
\end{array}
$$
{\bf Remark.} For another choice of the gauge function
$u(x)$, only the function $Q^\pm_m (x)$ differs by a
multiple of a certain rational function of $x^N$, which 
makes no effect as far as the Bethe equation is
concerned.
$\Box$ \par \vspace{0.2in} \noindent
\begin{lemma}\label{lem:eta} 
Let $q$ be a primitive $N$th
root of  unity,   $k, l$ be integers with
$q^k+ q^l \in \RZ$.  Then
$q^{k+l} = 1$. 
\end{lemma}
{\bf Proof.} By the odd property of $N$, 1 is
the only  real number among the
$N$th roots of unity. We may 
assume that  $1, q^k, q^l$ are three
distinct numbers . The following conditions are
equivalent,
$$
q^k+ q^l \in \RZ  \ \ 
\Longleftrightarrow \ 
q^k- q^{-l} \in \RZ \ \
\Longleftrightarrow \ 
q^{k+l}- 1 \in q^l \RZ \ .
$$
By interchanging $k$ and $l$ in the above relations, one
concludes that 
$(q^{k+l}- 1 ) \in q^k \RZ \cap
q^l
\RZ$.   By $q^k \neq \pm q^l$,  
$q^k \RZ
\cap
q^l \RZ$ consists of only the zero element. Hence
$q^{k+l} = 1$.
$\Box$ \par \vspace{0.2in} 

For the Bethe equation (\ref{QmEq}), one needs only to
consider the plus-part of the equation because
of the following result.
\begin{proposition} \label{prop:Qminus}
For $m \in \ZZ_N$, we have  $
 Q_m^-(x) = 0 $, $ |x\rangle_m^- = \vec{0}$, and 
\begin{eqnarray}
|x\rangle_m^+ = 2 q^{-m} |x\rangle_m^e u(qx)  = 2
|x\rangle_m^ou(x)  \ . \label{x+}
\end{eqnarray}
\end{proposition}
{\bf Proof.} Let $Q_m^- (x)$ be a
non-zero polynomial solution of  (\ref{QmEq}) for some
$\Lambda(x)$ with $\Lambda(0) = q^l+1$, and write
$Q_m^- (x) = x^r Q_m^{- *} (x)  $ with 
$Q_m^{- *} (0)  \neq 0$. By  comparing the
$x^r$-coefficients of {\rm (\ref{QmEq})}, we have 
$-q^{-m} \Lambda (0) Q_m^{- *} (0)=  ( q^{-r} + q^r )
Q_m^{- *}(0)$, hence 
$$
-q^{-m} (1+ q^l)=  q^{-r} + q^r  \ .
$$
By Lemma \ref{lem:eta},  $q^l= q^{2m}$, which implies 
$q^r = - q^{\pm m}$,  a contradiction to the odd property
of the integer $N$. Therefore the only solution   
$Q_m^-(x)$ for the negative-part of 
(\ref{QmEq}) is the trivial one. Since  
$Q_m^-(x)$ is of  the form 
$\langle  \varphi
|x\rangle_m^-$ for an eigenvector  $\langle \varphi|$  of
$T(x)$ in 
$\stackrel{L}{\otimes} \CZ^{N*}$, and all such
vectors
$\langle 
\varphi |$ form a basis of $\stackrel{L}{\otimes}
\CZ^{N*}$, hence 
$|x\rangle_m^- = \vec{0}$ for all
$m$. Then follows the  relation (\ref{x+}).
$\Box$ \par \vspace{0.2in} \noindent
We are going to derive some general properties of the 
solutions $Q^+_m (x)$ of (\ref{QmEq}). 
\begin{lemma}\label{lem:Qm}
For a  polynomial $\Lambda (x)$   with $\Lambda (0)=
q^l+1$, the necessary and sufficient condition of
$\Lambda(x)$ for the existence of a  non-zero polynomial
solution $Q_m^+(x)$ of the equation  {\rm (\ref{QmEq})}
with the eigenvalue $\Lambda(x)$
is given by the relation $q^l= q^{2m}$. In this
situation, with 
$Q^+_m(x) = 
x^r Q^{+ *}_m (x)$ and
$Q^{+ *}_m (0) \neq 0$, one has $q^r= q^{\pm m}$.
\end{lemma}
{\bf Proof. } Let
$Q_m^+(x)$ be a non-zero solution of  (\ref{QmEq}) and
write $Q_m^{+} (x) = x^r Q_m^{+ *} (x)  $ with 
$Q_m^{+ *} (0)  \neq 0$. By the same argument as in
Proposition \ref{prop:Qminus}, we have
$$
q^{-m} (1+ q^l)=  q^{-r} + q^r  \ ,
$$
hence $q^l= q^{2m}$ and $q^r= q^{\pm m}$ by 
Lemma \ref{lem:eta}. It remains to show the "sufficient"
part of the statement. For
$q^l= q^{2m}$,  by (\ref{formula}) (\ref{x+}), one
concludes
$|x\rangle_m^+$ can not be identically zero. We
claim that the solution $Q_m^+(x)$ in
(\ref{QmEq}) has non-trivial solutions.
Otherwise, this implies that
$|x\rangle_m^+$ is always the zero vector for all $x$ by
the same argument as in Proposition
\ref{prop:Qminus}, hence a contradiction. 
$\Box$ \par \vspace{0.2in} \noindent
\begin{proposition}\label{prop:Qplus}
Let $m$ be an integer between $0$ and $M$, and
$Q_m^+(x), Q_{N-m}^+(x)$ be solutions of 
$(\ref{QmEq})$ for  $m, N-m$ respectively which
arise from the evaluation of eigenvectors $\langle
\varphi|$ of
$T_{\vec{h}}(x)$ on the Baxter vector. Then 
$Q_m^+(x), Q_{N-m}^+(x)$ are elements in $x^m
\prod_{j=0}^{L-1} (1-x^Nc_j^N)
\CZ[x]$.
\end{proposition}
{\bf Proof.} \ The divisibility of $Q_m^+(x),
Q_{N-m}^+(x)$  by $x^m$ follows easily from Lemma
\ref{lem:Qm}, so only the factor
$\prod_{j=0}^{L-1} (1-x^Nc_j^N)$ remains to be verified.
As 
$Q_l^+(x)$ is of the form $\langle \varphi|x
\rangle_l^+$ for some vector $\langle \varphi|$ in
$\stackrel{L}{\otimes} \CZ^{N *}$, and by 
(\ref{formula}) (\ref{x+}), it suffices 
to show that the following divisibility of
polynomials,
\begin{eqnarray*}
\prod_{j=0}^{L-1} (xc_j ; \ w)_{M+1} &| \ u(qx) &
\prod_{j=0}^{L-1} 
\frac{(xc_j ; \ \omega^{-1})_{k_j+n+1} }{(xc_j ;
\ \omega )_{k_j+n+1} } \ , \\
\prod_{j=0}^{L-1} (xc_j\omega^{M+1} ; \ w)_M  & | \ u(x)&
\prod_{j=0}^{L-1} 
\frac{(xc_jq^{-1} ; \ \omega^{-1})_{k_j+n+1}
}{(xc_jq ; \ \omega )_{k_j+n+1} } \ .
\end{eqnarray*}
By the form of $u(x)$ in (\ref{udef})   and  the
relation $q
\omega^{M+1+j}= \omega^{j+1}$ for $j \in
\ZZ$, the above relations are easily seen. 
$\Box$ \par \vspace{0.2in} \noindent
For our purpose, the
functions $Q_m^+(x)$ we shall consider are only
those arising from eigenvectors of the transfer
matrix, hence $Q_m^+(x)$ is in the  form of
Proposition
\ref{prop:Qplus}. For the rest of this paper, the
letter $m$ will always denote an integer between
$0$ and $M$, 
$$
0 \leq m \leq M \ . 
$$
We shall conduct our discussion of the plus-part of the
equation (\ref{QmEq}) for the sectors
$m, N-m$ simultaneously by introducing the 
polynomials
$\Lambda_m(x), Q_m ( x)$ via the relation,
\be
(\Lambda_m(x) , \ x^m
\prod_{j=0}^{L-1} (1-x^Nc_j^N) Q_m ( x) ) = 
(q^{-m}\Lambda(x), \ Q_m^+(x) ) , \ \ (q^m
\Lambda(x) , \ Q_{N-m}^+(x) )  \ .
\ele(sub)
Then the relations (\ref{QmEq}) for both the $m$
and $ N-m$ sectors  are reduced to the following:
\be
 \Lambda_m (x) 
Q_m (x) = q^{-m} 
\prod_{j=0}^{L-1}(1-xc_jq^{-1}) 
Q_m (xq^{-1})+ q^m \prod_{j=0}^{L-1} (1+x c_j) 
 Q_m (xq) \ ,
\ele(rBeq)
where $Q_m(x), \Lambda_m(x)$ are polynomials with 
$$
\begin{array}{lll}
{\rm deg.}Q_m (x)
\leq   ML-m ,  &&\\
 {\rm deg.}
\Lambda_m (x) \leq 2[\frac{L}{2}], &
\Lambda_m (x)= \Lambda_m (-x), & \Lambda (0)= 
q^m + q^{-m} \ .
\end{array}
$$
The general mathematical problem will be  the
structure of the solution space of the
Bethe equation (\req(rBeq)) for a given positive integer
$L$. First we are going to derive the detailed answer of
the Bethe solutions for the simplest case $L=1$.  

${\bf L=1 .}$ We have
$\Lambda_m (x)=  q^m+q^{-m}$, and  ${\rm deg.}
Q_m(x) \leq M-m$.
\newtheorem{theorem}{Theorem}
\begin{theorem} \label{thm:L1Qm}
The solutions  
$Q_m(x)$s of ${\rm (\req(rBeq))}_{L=1}$ forms an
one-dimensional vector space  generated by the following 
polynomial of degree $M-m$,
\be
B_m(x) := 1+ \sum_{j \geq 1} (\prod_{i=1}^j 
\frac{q^{m+i-1} - q^{-m-i} }{q^m+q^{-m}-q^{-m-i}-q^{m+i}})
(xc_0)^j \ .
\ele(Bm)
(Note that  the coefficients in the above
expression are zero for $j > M-m$.)
\end{theorem}
{\bf Proof.} \ Write 
$$
Q_m(x) = \sum_{j=0}^{M-m} \beta_j  (
xc_0)^j \ .
$$
Then $(\req(rBeq))_{L=1}$ is equivalent to
the following system of equations of $\beta_j$s,
\bea(l)
(q^m+q^{-m}-q^{-m-j}-q^{m+j}) \beta_j =
(  q^{m+j-1} - q^{-m-j} ) \beta_{j-1}  \ , \ \ \ j \in
\ZZ_{\geq 0} \ , 
\elea(L1eq)
where $\beta_k$ is defined to be zero for 
$k$ not between $0$ and $M-m$. As the
values 
of 
$q^m+q^{-m}-q^{-m-j}-q^{m+j}$, $ q^{-m-j} - q^{m+j-1}
$ are all non-zero, the polynomial $Q_m(x)$ is
determined by $\beta_0$, (or
equivalently $\beta_{M-m}$), through the recursive
relations (\req(L1eq)). With $\beta_0=1$, this 
provides the basis element $B_m(x)$.
$\Box$ \par \vspace{0.2in} \noindent
\newtheorem{corollary}{Corollary}
\begin{corollary}\label{cor:L1}
The vector space of all 
polynomial solutions of ${\rm (\req(rBeq))}_{L=1}$,
(without the restriction of the degree of $Q_m(x)$), is
$\CZ[x^N] B_m(x)$.
\end{corollary}
{\bf Proof.} \ 
By using the fact 
$$
\begin{array}{lll}
q^{m+j-1}- q^{-m-j} = 0 &\Longleftrightarrow& j \equiv
M-m+1 \pmod{N} \ , \\
q^m+q^{-m}- q^{m+j} - q^{-m-j} = 0 &\Longleftrightarrow&
j
\equiv 0, \ N-2m \pmod{N} \ ,
\end{array}
$$
and the relation (\req(L1eq)) for those $j$ with 
$ M-m+1 + lN < j \leq (l+1) N, \ \ ( l \in \ZZ_{\geq0})
$, any solution $\sum_{k\geq 0} \beta_j (c_0x)^k $  of
${\rm (\req(rBeq))}_{L=1}$ must have 
$\beta_k=0$ except $k \equiv 0,  \ldots, M-m
\pmod{N}$,  hence is an element of $\CZ[x^N] B_m(x)$ by
Theorem 
\ref{thm:L1Qm}.
$\Box$ \par \vspace{0.2in} 
For the Bethe equation 
(\req(rBeq)) with 
$L >1$, by the scaling of the
variables, 
$$
x \mapsto \lambda^{-1} x \ , \ \ c_j \mapsto \lambda c_j
\  \ \ \ \ \  {\rm for} \ \ \lambda \in \CZ^* \ ,
$$ 
we may assume that the $x^j$-coefficients of  
polynomials, $\Lambda_m(x), Q_m(x)$, are always 
homogenous functions of $c_0, \ldots, c_{L-1}$ with the
degree $j$. As (\req(rBeq)) is invariant under 
permutations of $c_j$s, the coefficients of
the polynomials of $x$ involved in (\req(rBeq)) depend
only on the elementary symmetric functions of
$c_j$s,
$$
s_j = \sum_{ i_1<\ldots <i_j } c_{i_1} \cdots
c_{i_j} \ , \ \ \ \ {\rm for} \ j=1, \ldots, L \ .
$$
We shall denote
$$
 Q_m(x) = \sum_{j=0}^d \alpha_j x^j \ , \ \ \ d:= {\rm
deg.} Q_m(x) \ \ ( \leq LM-m)  \ ,
$$
and define $\alpha_j$ to be zero for 
$j$ not between $0$ and $d$. 
For the rest of this section, we shall only consider
the case, $L=2, 3$.

${\bf L=2 .}$ We have only two elementary symmetric
functions of $c_j$s: 
$$
s_1= c_0+c_1 \ , \ \ \ s_2= c_0c_1 \ . 
$$
\begin{lemma}\label{lem:det0}
Let $n$ be an odd positive integer, $A$ be a $n
\times n$-matrix with the complex entries $a_{i,j}$
satisfying the relations
$$
a_{i,j}= (-1)^{i+j+1} a_{n-j+1, n-i+1} \ , \ \ 
{\rm for } \ 1 \leq i, j \leq n \ .
$$
Then  $A$ is a degenerated matrix. 
\end{lemma}
{\bf Proof.} Write $n= 2h+1$. The determinant of
$A$ can be expressed by
\begin{eqnarray*}
{\rm det}(A)  & = &\sum_{\sigma} {\rm sgn}(\sigma)
\prod_{i=1}^n a_{i, \sigma(i)}   = \sum_{\sigma} {\rm
sgn} (\sigma)
\prod_{i=1}^n (-1)^{i+\sigma(i)+1} a_{n-\sigma(i)+1, 
n-i+1} \\ 
 & = & {\rm sgn}(\sigma_0) (-1)^{h+1}
\sum_{\sigma'} {\rm sgn} (\sigma') \prod_{j=1}^n 
a_{\sigma'(j),  j} = {\rm sgn}(\sigma_0) (-1)^{h+1}
{\rm det}(A) \ , 
\end{eqnarray*}
where the indices $\sigma, \sigma'$, run through all 
permutations of $\{1, \ldots, n \}$, and $\sigma_0$ is
the one  defined by 
$\sigma_0(i)=n-i+1$. By
${\rm sgn}(\sigma_0)=(-1)^h$, we have 
${\rm det}(A)=0$. 
$\Box$ \par \vspace{0.2in} \noindent
In the equation $(\req(rBeq))_{L=2}$,  $\Lambda_m(x)$ is
an even polynomial of degree $\leq 2$, and ${\rm deg.}
Q_m(x) \leq 2M-m$. By comparing the coefficients of the
highest degree of $x$ in (\req(rBeq)), we have
$$
\Lambda_m(x)= 
(q^{m+d}  +  q^{-m-d-2})x^2 s_2 +q^m+q^{-m} \ . 
$$
For $k \in \ZZ$, we define 
\bea(l)
v_k = q^k+q^{-k}- q^m - q^{-m}, \\ 
\delta_k = (q^{k-1}-q^{-k}) s_1, \\ 
u_k = (q^{k-2} + q^{-k}-
q^{m+d}-q^{-m-d-2} ) s_2 \ .
\elea(L2coef)
Then $(\req(rBeq))_{L=2}$ is  equivalent
to the  system of linear equations of $\alpha_j$s,
\bea(l)
v_{m+j} \alpha_j   + \delta_{m+j}  \alpha_{j-1}  
 + u_{m+j}  \alpha_{j-2} = 
0 \ , \ \ \ \  j \in \ZZ_{\geq 0}  \ .
\elea(L2eqn)
In fact, the 
non-trivial relations in the above system are those $j$
between $1$ and $d+1$, hence the matrix form of
(\req(L2eqn)) is given by
\bea(l)
\left( \begin{array}{cccccc}
\delta_{m+d+1} & u_{m+d+1} &0&\cdots & 0 & 0 \\
v_{m+d}& \delta_{m+d} &u_{m+d}&\cdots  & 0 &0 \\
0& \ddots&\ddots &\ddots  & &\vdots \\
\vdots& \ddots & \ddots & \ddots & \ddots &0\\
\vdots& \ddots &0 & v_{m+2} & \delta_{m+2} &u_{m+2}\\
0&  \cdots & & 0 & v_{m+1} & \delta_{m+1} 
\end{array} 
\right) 
\left( \begin{array}{c}
\alpha_d \\
\alpha_{d-1} \\
\vdots\\
\vdots\\
\vdots\\
\alpha_0
\end{array} 
\right) = \vec{0} \ .
\elea(Meqn2)
\begin{theorem} \label{thm:L2sol}
The equation ${\rm (\req(rBeq))}_{L=2}$ has a
non-trivial solution $Q_m(x)$ if and only if ${\rm deg.}
Q_m(x) = M -m +m'$ for $0 \leq m' \leq M$. For each such
$m'$, the eigenvalue $\Lambda_m(x)$ in ${\rm
(\req(rBeq))}_{L=2}$ is equal to $
\Lambda_{m,m'}(x) :=  q^{\frac{1}{2}} (q^{m'-1} + 
q^{-m'-2})x^2 s_2 +q^m+q^{-m}$, and the corresponding
solutions of $Q_m(x)$  form an one-dimensional space
generated by a
 polynomial $B_{m,m'}(x)$ of degree $M-m+m'$ with 
$B_{m,m'}(0) = 1$. ( Here $q^{\frac{1}{2}}:= q^{M+1}$. ) 
\end{theorem}
{\bf Proof.} Denote $d = {\rm deg.} Q_m(x) $. Among
those
$v_j$s of the entries of the square matrix 
(\req(Meqn2)), there is at most one zero term, which
is given by $v_{N-2m}= 0$. If
$Q_m(0) = 0$, this implies $Q_m (x)= x^{N-2m}Q_m^*(x) $
with 
$Q_m^*(0) \neq 0$, and each coefficient of the polynomial
$Q_m^*(x)$ is expressed by a $Q_m^*(0)$-multiple of a
certain polynomial of
$c_0, c_1$. By setting
$c_1=0$,
$x^{N-2m}Q_m^*(x)Q_m^*(0)^{-1} $ gives rise to a solution
of
${\rm (\req(rBeq))}_{L=1}$, which contradicts the
conclusion of Corollary 
\ref{cor:L1}. Therefore 
$Q_m(0) \neq 0$. If 
$d $ is less than $ M-m$, again by setting
$c_1=0$, the function $Q_m(x)$ gives rise to a solution
of 
${\rm (\req(rBeq))}_{L=1}$ of degree $< M-m$, a
contradiction to Theorem \ref{thm:L1Qm}. Hence $d=
M-m+m'$ for  $0 \leq m' \leq M$. It remains to
show for each such $m'$, the  solutions  $Q_m(x)$ 
 form an one-dimensional vector space. As any 
non-trivial solution $Q_m(x)$ must have $Q_m(0) \neq 0$, 
this implies the injectivity of the following linear
functional of the solution space, 
$$
Q_m(x) \ \mapsto Q_m(0) \in \CZ \ .
$$
So one needs only to show the existence of a 
non-trivial solution $Q_m(x)$, which is equivalent
to the degeneracy of the square matrix of size 
$d+1$ in the left hand side of (\req(Meqn2)). Write this
square matrix in the form
\bea(l)
\left( \begin{array}{cc}
A & 0 \\
C& B 
\end{array} 
\right)  
\elea(ABCL2)
where $ C$ is a  $(d-2m')\times (2m'+1)$-matrix, 
$A, B$ are the tri-diagonal square matrices of the 
size $2m'+1$, $ d-2m'$ respectively. The explicit
form of $A$ is given by 
$$
\begin{array}{l}
A = \left( \begin{array}{cccccc}
\delta_{M+1+m'} & u_{M+1+m'} &0&\cdots  & 0 &0 \\ 
v_{M+m'}& \delta_{M+m'}
&u_{M+m'}&0&\ddots    &0 \\ 
0&  v_{M+m'-1}& \delta_{M+m'-1}
&u_{M+m'-1} & \ddots &\vdots \\
\vdots& \ddots & \ddots & \ddots & \ddots &0\\
\vdots& \ddots & \ddots & \ddots & \ddots &0\\
\vdots& \ddots &0 & v_{M+2-m'}  & \delta_{M+2-m'} &
u_{M+2-m'}\\
0&  \cdots & & 0 & v_{M+1-m'} &
\delta_{M+1-m'} 
\end{array} 
\right).
\end{array}
$$
By (\req(L2coef)), we have $
v_j = v_{N-j}, \delta_j = - \delta_{N+1-j} , u_j =
u_{N+2-j}$. This implies that the matrix $A$ satisfies
the condition of Lemma \ref{lem:det0}, hence ${\rm
det}(A) = 0$.   Therefore the square matrix
(\req(ABCL2)) has the zero determinant.
$\Box$ \par \vspace{0.2in} \noindent
{\bf Remark.}
By the above theorem,
the following data are in one-one correspondence with
integers 
$m, m'$ between $0$ and $M$,
$$
(m , m') \ \longleftrightarrow \ \Lambda_{m,m'}(x) \ 
\longleftrightarrow B_{m,m'}(x) \ .
$$
The characterization of $B_{m,m'}(x)$ is given as 
the (unique) polynomial with ${\rm deg.}
B_{m,m'}(x)= M-m+m'$ and $B_{m,m'}(0)=1$, whose 
coefficients $\alpha_j$s satisfy the equation 
(\req(L2eqn)) with 
\bea(ll)
v_k =& q^k+q^{-k}-q^m-q^{-m},  \\
\delta_k =& (q^{k-1}-q^{-k}) s_1, \\ 
u_k= & (q^{k-2}+q^{-k}-
q^{m'-\frac{1}{2}}-q^{-m'-\frac{5}{2}})s_2 . 
\elea(L2coefm')
$\Box$ \par \vspace{0.2in} 

${\bf L=3 .}$ There are three elementary
symmetric functions of $c_j$s,
$$
s_1 = c_0+c_1+c_2 \ , \ \ 
s_2 =c_0c_1+c_1c_2+c_2c_0 \ , \ \ 
s_3= c_0c_1c_3 \ .
$$
A non-trivial
solution  $Q_m(x)$ of ${\rm (\req(rBeq))}_{L=3}$ has the
degree
$d  \leq 3M-m $, and $\Lambda_m(x)$ is of the form 
\be
\Lambda_m(x) = \lambda_m x^2 + q^m + q^{-m} \ .
\ele(LmL3)
Note that $\lambda_m$ is a homogeneous function of
$c_j$s of degree 2. For $k \in
\ZZ$, we define
\begin{eqnarray*}
  w_k&=& q^k +q^{-k}-q^m-q^{-m}, \\
v_k &= &(q^{k-1} -q^{-k})s_1, \\ 
\delta_k& = &(q^{k-2} +q^{-k})
s_2 , \\ 
u_k&= &(q^{k-3}-q^{-k})s_3  \ .
\end{eqnarray*}
The equation
${\rm (\req(rBeq))}_{L=3}$ is equivalent to the system of
linear equations of $\alpha_j$s,
\bea(l)
w_{m+j} \alpha_j +v_{m+j} \alpha_{j-1}+ 
(\delta_{m+j}-\lambda_m)
\alpha_{j-2} + u_{m+j} \alpha_{j-3} = 0 \ , 
\ \ j \in \ZZ_{\geq 0} \ .
\elea(L3eq)
The non-trivial relations of the above system are 
those for
$j$ between $1$ and $d+3$.
\begin{lemma}\label{lem:L3Qm}
Let $Q_m(x)$ be a non-trivial polynomial solution of
${\rm (\req(rBeq))}_{L=3}$ for some $\Lambda_m(x)$.
Then  the degree of $Q_m(x)$ is equal to
$3M-m$ with $Q_m(0) \neq 0$. 
\end{lemma}
{\bf Proof.} \ First we note that for $j$ between $m+1$
and
$3M$, the only  possible 
$w_j$s with zero
value are given by 
$$
w_{N-m} = w_{m+N} = 0 \ .
$$
Let $r$ be the zero multiplicity of $Q_m(x)$ at $x=0$. 
If $Q_m(0) = 0$, by (\req(L3eq)) we have $r= N-2m$ or
$N$. The polynomial 
$\alpha_r^{-1}Q_m(x)$  with $c_1=c_2=0$ is a
non-trivial solution of ${\rm (\req(rBeq))}_{L=1}$ with
the zero multiplicity $r$. By Corollary
\ref{cor:L1},  $r$ must be equal to $N$ and the
degree of $Q_m(x)$ is at least $N+M-m$, which
contradicts our assumption,
$d \leq 3M-m$. Therefore $Q_m(0) \neq 0$.
By the relation of $j=d+3$ in (\req(L3eq)), 
we have  $q^{m+d} = q^{-m-d-3}$, hence the only
possible values of 
$d$ are $M-m-1$ , $3M-m$. If $d=M-m-1$, by $w_j \neq 0$ 
for $m+1 \leq j <M$,  the function $Q_m(x)$ with
$c_1=c_2=0$ gives rise to a non-trivial solution of ${\rm
(\req(rBeq))}_{L=1}$ with the degree 
$< M-m$, a contradiction to Theorem 
\ref{thm:L1Qm}. Therefore $d=3M-m$. 
$\Box$ \par \vspace{0.2in} \noindent 
By the above lemma, the "eigenfunction" $Q_m(x)$ for an
eigenvalue $\Lambda_m(x)$ is unique up to a non-zero
constant, hence there is an one-one correspondence
between the eigenvalues and eigenstates of the Bethe
equation $(\req(L3eq))_{L=3}$ for a given
$m$.  By $d=
3M-m$, the $(d+3)$-th relation in the system
(\req(L3eq)) is redundant,
hence the matrix form of (\req(L3eq)) becomes
\bea(l)
\left( \begin{array}{cc}
A-\lambda_m & 0 \\
C& B 
\end{array} 
\right) \left( \begin{array}{c}
\alpha_d \\
\vdots\\
\vdots\\
\alpha_0
\end{array} 
\right) = \vec{0} \ , \ \ \ \ d=3M-m \ , 
\elea(ABCL3)
where $A $ is the $N \times N$ matrix,
\bea(l)
A=  \left( \begin{array}{ccccccc}
\delta_{N-1}' & u_{N-1}' &0&\cdots  &  &0 &0
\\ v_{N-2}'& \delta_{N-2}' 
&u_{N-2}'&0&\ddots  &  &\vdots \\
w_{N-3}'&v_{N-3}'& \delta_{N-3}' 
&u_{N-3}'&\ddots &  & 0  \\
0& \ddots&\ddots &\ddots  & \ddots & &\vdots \\
\vdots& \ddots & \ddots & \ddots & \ddots &\ddots  &0\\
\vdots& \ddots &0&w_1' & v_1'  & 
\delta_1' &u_1'\\
0&  \cdots&  & 0 & w_0' & v_0' &
\delta_0'
\end{array} 
\right) 
\elea(AL3)
with the entries defined by
$$
\begin{array}{ll}
w_k'= q^{k+\frac{3}{2}}
+q^{-k-\frac{3}{2}}-q^m-q^{-m}, &  
v_k' =
(q^{k+\frac{1}{2}} -q^{-k-\frac{3}{2}})s_1, \\ 
\delta_k' = (q^{k-\frac{1}{2}}
+q^{-k-\frac{3}{2}}) s_2 , &
u_k'=
(q^{k-\frac{3}{2}}-q^{-k-\frac{3}{2}})s_3  \ , 
\end{array}
$$
and
$B, C$ are the following matrices,
$$  
\begin{array}{l}
B =  \left( \begin{array}{cccccc}
\delta_{M+1}-\lambda_m & u_{M+1} &0&\cdots  & 0 & 0 \\
v_{M}& \delta_{M}-\lambda_m 
&u_{M}&0&\ddots    &\vdots \\
w_{M-1}&v_{M-1}& \delta_{M-1}-\lambda_m 
&u_{M-1}&\ddots   & 0  \\
0& \ddots&\ddots &\ddots  & \ddots &\vdots \\
\vdots& \ddots & \ddots & \ddots & \ddots &0\\
\vdots& \ddots &w_{m+3} & v_{m+3}  & 
\delta_{m+3}-\lambda_m &u_{m+3}\\
0&  \cdots & 0 & w_{m+2} & v_{m+2} & 
\delta_{m+2} -\lambda_m\\
0&  \cdots & 0 & 0 & w_{m+1} & 
v_{m+1} 
\end{array} 
\right), \\ 
C =  \left( \begin{array}{cccccc}
0&\cdots& \cdots &0 &w_{M+1} & 0 
\\
0&\cdots  &&& 0 & w_{M}  
 \\
\vdots& \cdots&\cdots &\cdots  & \cdots &0 \\
0& \cdots &  & \cdots & \cdots &0
\end{array} 
\right) \ , \ \ w_M=w_{M+1}= q^{\frac{1}{2}}+
q^{\frac{-1}{2}} -q^m- q^{-m} \ . 
\end{array}
$$
Note that the coefficient-matrix of the equation 
(\req(ABCL3)) is of the size
$(d+2)
\times (d+1)$, while there are only $d+1$ variables
$\alpha_j$s to be solved. The matrix $B$ is  
equal to the upper-left $(M-m+1) \times (M-m)$ submatrix
of the square matrix $A-\lambda_m I$, of which the 
entries
$a_{ij}$, 
$1 \leq i, j \leq N$, satisfy the relations 
$a_{i,j}= (-1)^{i+j} a_{N-j+1, N-i+1}$.
\begin{theorem}\label{thm:L3sol}
For $0 \leq m \leq M$, the condition of the eigenvalue  
$\Lambda_m(x)= \lambda_m x^2 +
q^m + q^{-m} $ with a non-trivial solution $Q_m(x)$ of
the equation
${\rm (\req(rBeq))}_{L=3}$ is determined by the solution
of
${\rm det}(A-\lambda_m) =0$, where $A$ is the matrix
defined by
$(\req(AL3))$. For
each such 
$\Lambda_m(x)$, there exists a unique (up to
constants) non-trivial polynomial solution 
$Q_m(x)$ of 
${\rm (\req(rBeq))}_{L=3}$, and the degree $Q_m(x)$
is equal to $3M-m$ with  $Q_m(0) \neq 0$.
\end{theorem} 
{\bf Proof.} \ By Lemma \ref{lem:L3Qm}, one needs only
to show the existence of a non-trivial solution
$\alpha_j$s of (\req(ABCL3)) for
a    
$\lambda_m$ with  ${\rm
det}(A-\lambda_m) =0$. For $m=M$, it is obvious as there
is no matrix
$B$,  and $C$ is  zero. For
$m <M$, with a given $\lambda_m$, there exists a
non-trivial vector in the kernel of 
$A-\lambda_m $, 
$$
( A -\lambda_m )\left( \begin{array}{c}
\alpha_{3M-m} \\
\vdots\\
\vdots\\
\alpha_{M-m} \end{array} 
\right) = \vec{0} \ .
$$
As the $u_j$s appearing in the matrix $B$ are all 
non-zero, by the fact that the rank of $B$ is at most
$M-m$, one can extend the above $\alpha_j \ (M-m \leq j
\leq 3M-m)$ to a  solution $\alpha_j$s of (\req(ABCL3)).
The result then follows. 
$\Box$ \par \vspace{0.2in} \noindent
{\bf Remark.} By the above theorem, the
eigenstate $Q_m(x)$ is unique for a
given $\Lambda_m(x)$. It implies that for each
$m$, the eigenvalues $\Lambda_m(x)$ and eigenstates
$Q_m(x)$ of the Bethe equation $(\req(rBeq))_{L=3}$ are
in one-one correspondence. Note that these
$Q_m(x)$ are obtained under the constraint of
$\Lambda_m(x)$  with the form (\req(LmL3)), a conclusion
by the analysis of the transfer matrix in Proposition
\ref{prop:Lambda}, which we will refer as the "physical"
criterion while comparing the usual
Bethe-ansatz-technique in the discussion of the next
section.
$\Box$ \par \vspace{0.2in} \noindent

\section{The Degeneracy and Physical Solution Discussion 
of the Bethe Ansatz Relation} 
In this section, we first discuss the
degeneracy relation of eigenspaces of the transform
matrix
$T(x)$ in
$\stackrel{L}{\otimes}\CZ^{N*}$ with respect to the
Bethe solutions we obtained in Sect. 5. As
before, 
$\Lambda (x)$ denotes the eigenvalues of $T(x)$. 
By Proposition \ref{prop:Lambda}, the constant term of
$T(x)$ is given by
$$
T_0  = D + 1 , \ \ \ \ \ D := q^{-L}
\stackrel{L}{\otimes} Y \ ,
$$
with the eigenvalue $\Lambda(0)$, which is  of  the
form
$q^l+1$. For $l \in \ZZ_N$, we denote 
$$
\EZ_L^l = {\rm the \ eigensubspace \ of \ }
\stackrel{L}{\otimes}\CZ^{N*} \ {\rm of \ 
the \ operator \ }
D  {\rm \ with \ the \ eigenvalue \ } q^l \ ,
$$
which is of the 
dimension $N^{L-1}$. 
By Lemma
\ref{lem:Qm}, for $0 \leq m \leq M$, the equation
(\req(rBeq)) describes the relation of   
$\Lambda(x)$ and $Q_*^+(x)$ through (\req(sub)) when
$\Lambda(0)= q^{2m}+1$ or $ q^{2(N-m)}+1$.
For $L=1$, $T(x)$ is the constant 
$q^{-1}Y+I$, and $\EZ_1^l $ is the eigenspace of
$Y$ for the eigenvalue
$q^{l+1}$. By the evaluation at the Baxter vector 
$|x\rangle_{m}^+$, $|x\rangle_{N- m}^+$ respectively,
both the spaces, $\EZ_1^m$ and 
$\EZ_1^{N-m}$, give rise to the same functional space
generated by 
$x^m(1-x^Nc^N) B_m(x)$  with 
$B_m(x)$ in Theorem
\ref{thm:L1Qm}. For $L=2, 3$, the expression (\req(T02))
of $T_2$ becomes
\bea(lll)
L=2 , & T_2&= q^{-1} c_0c_1 ( X\otimes Z + Z
\otimes X )
\ ,
\\ L=3 , & T_2 & = q^{-2}( c_0c_1  X\otimes Z \otimes Y +
 c_1c_2 Y \otimes  X \otimes Z+ c_0c_2 Z
\otimes Y
\otimes X )  \\
 &  & + \ q^{-1} ( c_0c_1 Z\otimes X \otimes I +
c_1c_2 I \otimes  Z \otimes X +  c_0c_2 X
\otimes I
\otimes Z ) \ .
\elea(ratT02)
We shall denote ${\cal O}_L$ the operator
algebra generated by the $L$-tensors of 
$X, Y, Z, I$
appeared in the corresponding expression of $T_2$. Then 
${\cal O}_L$ 
commutes with 
$D$, hence one obtains a ${\cal O}_L$-representation on
$\EZ_L^l$ for each $l$. 

${\bf L=2 .}$ The 
${\cal O}_2$ is a commutative algebra with the
generators
$X\otimes Z$,$Z \otimes X$, and it contains the
element 
$D ( = ZX \otimes XZ ) $. 
The ${\cal O}_2$-representation  on
$\stackrel{2}{\otimes}
\CZ^{N *}$ has the eigenspace decomposition 
indexed by the $( X \otimes
Z, Z \otimes X)$-eigenvalue $( q^i, q^j)$, 
or equivalently, the 
$(D, Z \otimes X)$-eigenvalue $(q^l, q^j)$,  where 
$i, j, l$ are elements in $\ZZ_N$ with the relation 
$q^l= q^{j+i}$. In fact, the eigenspace is one
dimensional with the basis 
$\langle \phi_{j_0, j_1}|$ defined by 
$$
\langle \phi_{j_0, j_1}| := \frac{1}{N^2} \sum_{k, k'
\in \ZZ_N}
\omega^{j_0k+j_1k'-kk'}\langle k,  k'| \ ,
$$
where $(j_0,
j_1)$ is related to $(i, j)$ by $(\omega^{j_0},
\omega^{j_1})= (q^i, q^j)$. 
The vectors
$\langle \phi_{j_0, j_1}|$, with
$\omega^{j_0+j_1}= q^l$, form a basis of
$\EZ_2^l$.  The permutation of tensor factors 
of  
$\stackrel{2}{
\otimes} \CZ^{N *}$ induces an automorphism of $\EZ_2^l$,
which interchanges the vectors  $\langle
\phi_{j_0, j_1}|$ and
$\langle \phi_{j_1, j_0}|$. The
eigenvalues of $T_2$ are given by  $c_0c_1(q^{l-j-1} +
q^{j-1})$ for $j \in \ZZ_N$, and the
corresponding eigenspace is generated by $\langle
\phi_{j_0, j_1}|$ and  
$\langle \phi_{j_1, j_0}|$, where $\omega^{j_1}= q^j,
\omega^{j_0+j_1}= q^l$, with the dimension equal to  
two for all $j$ except  
$j= (M+1)l$. By (\req(sub)), the index
$(l, j)$ which corresponds to $(m, m')$ of 
Theorem
\ref{thm:L2sol} is given by the relation,
$$
(q^l, q^j ) = (q^{2m}, q^{m-m'-\frac{1}{2}}) \ , \
(q^{-2m}, q^{-m-m'- \frac{1}{2}} ) \ .
$$
With the evaluation at the corresponding Baxter vector, 
$\langle \phi_{j_0, j_1}|$ and $\langle \phi_{j_1,
j_0}|$  give rise to the same eigenstate   
$B_{m,m'}(x)$ in  Theorem \ref{thm:L2sol}.

${\bf L= 3 .}$ We have 
\be
qD = (Z\otimes X \otimes I)( X
\otimes I \otimes Z)( I
\otimes Z \otimes X) \ .
\ele(3D)
With the identification, 
\begin{eqnarray}
U  = D^{-1/2}  Z\otimes X \otimes I , \ \ \ V
=  D^{-1/2}  X
\otimes I \otimes Z  \ , \ \label{UV}
\end{eqnarray}
${\cal O}_3$ is generated by $U, V$ which satisfy the
Weyl relation
$UV= q^2VU$ and   the $N$th power identity. Hence 
${\cal O}_3$ is the Heisenberg algebra and contains 
$D$ as a central element.   By (\req(ratT02)), 
$qD^{\frac{-1}{2}}T_2$ is the following
Hofstadter-like Hamiltonion, 
\bea(l)
c_0c_1 (   U  +
U^{-1} )
 +  c_0c_2 ( V +
V^{-1} ) +  c_1c_2 
( q 
D^{5/2}  UV + q^{-1}  D^{-5/2}  V^{-1}U^{-1} ) \ .
\elea(GSHof)
Note that the above Hamiltonian is a special
case of the Faddeev-Kashaev  Hamiltonian $H_{FK}$
with
$W= q^{-1}  D^{-5/2}  V^{-1}U^{-1} $,
$\alpha=\beta=
\gamma = 1$. Our conclusion on the sector $m=M$ is
equivalent to that in
\cite{FK} as it will become clearer later on. It is
known that there is a unique (up to  equivalence) 
non-trivial irreducible representation of ${\cal O}_3$,
denoted by
$\CZ^N_{\rho}$, which is
of the dimension $N$. For each $l$, 
$\EZ_3^l$ is equivalent to $N$-copies of
$\CZ^N_{\rho}$ as ${\cal
O}_3$-modules:  
$\EZ_3^l
\simeq N
\CZ^N_{\rho}$. For $0 \leq m \leq M$, we consider the
space
$\EZ_3^l$ with 
$q^l= q^{\pm 2m}$. The evaluation of $\EZ_3^l$
on $|x\rangle_{\pm m}^+$ gives rise to a $N$-dimensional
kernel in
$\EZ_3^l$. By Theorem
\ref{thm:L3sol}, there are $N$ 
polynomial solutions $Q_m(x)$ of degree $3M-m$ 
of 
${\rm (\req(rBeq))}_{L=3}$  for each of 
$N$ distinct eigenvalues $\Lambda_m(x)$. The $N$-dimensional
vector space spanned by those 
$Q_m(x)$s realizes the
irreducible  representation
$\CZ^N_{\rho}$ of the Heisenberg algebra
${\cal O}_3$.

Now we discuss the relation between the Bethe
equation (\req(rBeq)) and the usual  Bethe
ansatz formulation in literature. For the physical
interest, we are going to focus our attention only on the
case $L=3$. For 
$0
\leq m \leq M$, a solution $Q_m(x)$ in
$(\req(rBeq))_{L=3}$ must have $Q_m(0) \neq
0$ by Theorem \ref{thm:L3sol}. We write 
\be
\alpha_{3M-m}^{-1} Q_m (x) = \prod_{l=1}^{3M-m} ( x -
\frac{1}{z_l} ) \ , \ \ \ 
\  z_l \in \CZ^*  \ .
\ele(NormQ)
By setting $x= z_l^{-1}$ in 
${\rm (\req(rBeq))}_{L=3}$ , we obtain the
following relation among
$z_l$s, called the Bethe
ansatz equation in
the physical literature as in the case of usual
integrable Hamiltonian chains,
\begin{eqnarray}
 q^{m+\frac{3}{2}}  \prod_{j=0}^2 \frac{z_l +
 c_j}{qz_l - c_j} =   \prod_{n=1, n \neq l}^{3M-m}
\frac{ qz_l - z_n }{ 
z_l -  q z_n  } \ , & \ \ 1 \leq l \leq 3M-m \ .
\label{BA}
\end{eqnarray}
The following lemma is obvious.
\begin{lemma}\label{lem:BA}
For a polynomial $Q_m(x)$ of the form $(\req(NormQ))$, 
the Bethe ansatz relation $(\ref{BA})$ for 
the roots of  $Q_m(x)$ is equivalent
to the  divisibility of  
$q^{-m} 
\prod_{j=0}^2(1-xc_jq^{-1}) 
Q_m (xq^{-1})+ q^m \prod_{j=0}^2 (1+x c_j) 
 Q_m (xq)$ by $Q_m (x)$. In this situation, the
quotient  polynomial $\Gamma (x)$ of the latter pair has
the degree  at most $2$ with $\Gamma (0)=q^{-m}+q^m$.
\end{lemma}
As mentioned in the remark of Theorem \ref{thm:L3sol}, 
the eigenvalue of the Bethe equation
$(\req(rBeq))_{L=3}$ is described by the eigenstate
$Q_m(x)$, which is a polynomial with the
pre-described  zeros satisfying (\ref{BA}).
However, the quotient polynomial $\Gamma (x)$ in
Lemma \ref{lem:BA}   arising from  a solution of
(\ref{BA}) might not be an eigenvalue 
$\Lambda_m(x)$ with the form (\req(LmL3)) of the
Bethe equation 
$(\req(rBeq))_{L=3}$,
i.e., $\Gamma (x)$ might not be necessarily an even
function. A solution of (\ref{BA}) with a non-even
quotient polynomial 
$\Gamma(x)$ will be called a "non-physical" one.
For the sector $m=M$, there is no non-physical Bethe
ansatz solution of (\ref{BA}) by the following lemma.
\begin{lemma}\label{lem:BAm=M}
For $m=M$, the quotient polynomial $\Gamma (x)$ in 
Lemma $\ref{lem:BA}$ associated to a solution of 
$(\ref{BA})$ is always of the form $\Lambda_M(x)$ in
$(\req(LmL3))$.
\end{lemma}
{\bf Proof.} By Lemma \ref{lem:BA}, 
$\Gamma (x) = \sum_{j=0}^2 \gamma_j x^j$ and
$\gamma_0= q^{-M} + q^M$. It suffices to show the
vanishing of 
$\gamma_1$. We write 
$\alpha_{2M}^{-1} Q_M(x)= x^{2M} + \sum_{j=0}^{2M-1}
\beta_j x^j $ with $\beta_0 \neq  0$.  By comparing the 
$x$-coefficients of $
(\req(rBeq))_{L=3}$,  one has
$$
\gamma_1 \beta_0 + \gamma_0 \beta_1 = 
(-q^{-M-1} + q^M ) s_1 
\beta_0 +  (q^{-M-1} + q^{M +1})
\beta_1  \ , 
$$
which implies $\gamma_1=0$. 
$\Box$ \par \vspace{0.2in} \noindent 
With a further argument following the proof of the
above lemma, one can obtain the relations  (5.26),
(5.27)\footnote{The
$M$ in our paper is denoted by  $P$ in \cite{FK}. } 
of \cite{FK},  where the conclusion applies only
on the $M$-sector. In fact, 
the comparison of the $x^2$-coefficients of ${\rm
(\req(rBeq))}_{L=3}$ yields the relation,
$$
\lambda_M
= ( q^{M-1}+q^M)s_2 
+  (q^{M+1} - q^{M-1}  ) s_1 
\beta_1 \beta_0^{-1} + 
(q^{M-1} + q^{M+2} - q^M -q^{M+1}) 
\beta_2 \beta_0^{-1} \ , 
$$
an equivalent expression is the following one,
\begin{eqnarray}
 \lambda_M 
= ( q^{\frac{-1}{2}}+ q^{\frac{-3}{2}} )s_2 
+  (q^{\frac{1}{2}} - q^{\frac{-3}{2}}) s_1
\sum_{n=1}^{2M} z_n
 + 
(q^{\frac{3}{2}} + q^{\frac{-3}{2}}  - q^{\frac{1}{2}}
-q^{\frac{-1}{2}} ) 
\sum_{l<n} z_lz_n  \ . 
\label{lambda}
\end{eqnarray}
With the substitution, $
\mu = q^{\frac{1}{2}}c_0^{-1}, 
\nu = q^{\frac{1}{2}}c_1^{-1},  
\rho= q^{\frac{1}{2}}c_2^{-1} $, the expressions $(\ref
{BA})_{m=M}$, $(\ref{lambda})$ coincide with  (5.26)
(5.27) in  \cite{FK}. By Lemma \ref{lem:BAm=M}, the
Bethe ansatz relation (\ref{BA}) has been shown to be
equivalent to the Bethe equation $(\req(rBeq))_{L=3}$
for the $M$-sector. However, the parallel statement is
not true for the sector 
$m = M-1$, in fact, there does exist some "non-physical"
 Bethe ansatz solutions of (\ref{BA}). An 
obvious example is given by the  following one. By
$3M-m= N$, the collection of inverse of roots of $x^N-
\beta =0,
\ (\beta \neq 0)$, forms a solution of the Bethe ansatz
equation (\ref{BA}), and its associated quotient
polynomial $\Gamma (x)$ in Lemma \ref{lem:BA} is given by
$$
\Gamma (x) =  (q^{\frac{-3}{2}}
+ q^{\frac{3}{2}}) + (
 q^{\frac{-3}{2}}-q^{\frac{1}{2}})s_1 x   + 
(q^{\frac{-3}{2}} +q^{\frac{-1}{2}} )s_2 x^2 \ ,
$$ 
which is not an even polynomial, required by the
solutions of the equation (\req(rBeq)). This
shows that the inverse of roots of $x^N-
\beta =0$ for $\beta \neq 0$ provides a "non-physical" solution of Bethe ansatz
equation (\ref{BA}).  
By the above example, the constraint (\req(LmL3)) on
the eigenvalue $\Lambda_m(x)$  should be taken into
account in the discussion of Bethe ansatz solutions
of (\ref{BA}) for
 the spectrum problem of
 the transfer matrix in an arbitrary sector. Such a
consideration will become more crucial when the problem
of thermodynamic flux limit procedure is involved  in
the next section.

\section{Rational Bethe Equation for a generic $q$ }
In this section, we are going to study the rational Bethe
equation for a generic $q$. In particular for $|q|=1$,
it  is the infinity flux limit discussion, i.e. $N
\rightarrow
\infty $, of the models appeared in the last two
sections. In the discussion of this section, $q$ will
always mean a generic one. By using the formula  
(\req(sRep)), one has the canonical representation on
$\CZ^{\infty}$ of the Weyl algebra  generated by $Z, X$
with the relation
$$
ZX = q^2 XZ \ , \ \ \ \ Y := ZX \ .
$$
With $\vec{h} \in (\PZ^3)^L$, one defines the
transfer matrix $T_{\vec{h}}(x)$ as in the finite $N$
case, and then discusses  its spectrum $\Lambda (x)$.
Under the degenerating assumption (\req(rataspm)),
the method of the previous three sections can
equally be applied to reduce  the diagonalization
problem of the transfer matrix to the Bethe
solutions of the equation (\req(rBeq)). It is
important to notice that the form of (\req(rBeq))
is valid  for all finite
$N$ while keeping the size $L$ fixed. This enforces
us to use the same form  of the Bethe equation for a
generic $q$ as the one in  the finite
$N$ case, so that the formulation becomes
compatible with the infinity
$N$ limiting process.   An eigenvalue of the transfer
matrix, the same for the Bethe equation,
should be the generic analogy of
$\Lambda(x)$ appeared in the first relation of 
(\req(sub)), now denoted by $
\widetilde{\Lambda}_m (x) = q^m \Lambda_m (x) $; 
however, for the eigenstate of the Bethe
equation,  we will keep the infinity $N$ version
of 
$Q_m(x)$ in  (\req(rBeq)) .
The Bethe equation for a generic $q$ for $ m \in
\ZZ_{\geq0} $ is now described by
\be
\widetilde{\Lambda}_m (x) 
Q_m (x) =  
\prod_{j=0}^{L-1}(1-xc_jq^{-1}) 
Q_m (xq^{-1})+ q^{2m} \prod_{j=0}^{L-1} (1+x c_j) 
 Q_m (xq) \ ,
\ele(rBeqG)
where $\widetilde{\Lambda}_m(x)$ is an even polynomial
of degree $\leq
2[\frac{L}{2}]$ with 
$\widetilde{\Lambda}_m (0)= 
q^{2m}+1$, and $ Q_m(x)$ is a formal (power) series of
$x$, 
i.e., 
$$
Q_m(x) =\sum_{j \geq 0} \alpha_j x^j \ , \in
\CZ[[x]] \ . 
$$
As in the discussion of the last two sections, we
are going to make a similar analysis on solutions
of (\req(rBeqG)) for
$L
\leq 3$. For $L=1$, $\widetilde{\Lambda}_m (x)= 
q^{2m}+1$, and $B_m(x)$ in Theorem
\ref{thm:L1Qm} defines a formal series, which
becomes the basis of the solution space of
$(\req(rBeqG))_{L=1}$. Note that the description
of $\widetilde{\Lambda}_m (x)$ is consistent with
the spectrum of the operator $Y$.  For $L=2,3$,
by the expression (\req(ratT02)) of $T_2$, we
have the operator algebra 
${\cal O}_{L}$ as before, in which  $T_0$
is  a central element. 

${\bf L=2 .}$ The equation $(\req(rBeqG))_{L=2}$
is equivalent to the linear systems (\req(L2eqn)) of
$\alpha_j$s, where the coefficients of the
equations are defined by (\req(L2coefm')). By
Theorem
\ref{thm:L2sol}, the eigenvalue 
$\widetilde{\Lambda}_m(x)$ is given by 
$$
\widetilde{\Lambda}_{m,m'}(x) =  
(q^{m+m'-\frac{1}{2}} + 
q^{m-m'-\frac{5}{2}})x^2 s_2 +q^{2m}+ 1 \ , \ \ \ 
m' \in \ZZ_{\geq 0} \ .
$$
For $m, m' \in \ZZ_{\geq 0}$, the solutions
$Q_m(x)$ of $(\req(rBeqG))_{L=2}$ form an
one-dimensional space generated by an element
$B_{m, m'}(x) \in \CZ[[x]]$ with 
$B_{m,m'}(0)=1$. Note that the algebra  ${\cal O}_2$ is 
commutative, and the spectra of the ${\cal
O}_2$-representation 
$\stackrel{2}{\otimes}\CZ^{\infty *}$ give rise to the
eigenvalues of $T(x)$, which coincide with the above 
$\widetilde{\Lambda}_{m,m'}(x)$ for $m, m' \in
\ZZ_{\geq 0}$.

${\bf L= 3.}$ We consider the
operator $qD^{\frac{-1}{2}}T_2$, which is the Hamiltonian
(\req(GSHof)). For the sector $m$, its eigenvalue is
given by  
$q^{1-m} \widetilde{\lambda}_m$, where
$\widetilde{\lambda}_m$ is related to the  polynomial
$\widetilde{\Lambda}_m(x)$ by
$$
\widetilde{\Lambda}_m(x) =
\widetilde{\lambda}_m x^2 + q^{2m} + 1 \ .
$$
With
$\lambda_m:= q^{-m}\widetilde{\lambda}_m$, 
the $\lambda_m$s form the spectra 
of the quadric-diagonal square matrix $A$ (\req(AL3)) for a
generic $q$, which is now of
the infinity size as $M $ increases to  $\infty$.
For each  $\widetilde{\lambda}_m$, the
equation
$(\req(rBeqG))_{L=3}$ of $Q(x)$ is equivalent to  the
system (\req(L3eq)). It is easy to
see that there exists a unique solution
$Q_m(x)$ (up to a constant) by the generic property of
$q$.  Note that the same conclusion of $Q_m(x)$ equally
holds for an arbitrary  complex number
$\widetilde{\lambda}_m$, including those not in 
the spectra of $A$. This means that solutions of the
equation
$(\req(rBeqG))_{L=3}$ alone contain, but not
sufficiently determine  the eigenvalues
of the transfer matrix $T(x)$. Those
$\widetilde{\lambda}_m$s not from the spectra of 
$A$ correspond to the non-physical Bethe ansatz
solutions of the finite $N$ case, as 
discussed  in the last section.

\section{High Genus Curves and the Hofstadter Model}
We are now going back to the general situation in
Sect. 3. Note that the values $\xi_j^N$s of the curve
${\cal C}_{\vec{h}}$ in (\ref{eq:Cvh}) are determined
by $\xi_0^N$ and $x^N$. Denote
$$
y = x^N \ , \ \ \eta =\xi_0^N \ .
$$
The variables $(y, \eta)$ defines the curve 
\bea(l)
{\cal B}_{\vec{h}} : \ \ C_{\vec{h}}(y) \eta^2 +
(A_{\vec{h}}(y)-  D_{\vec{h}}(y))\eta 
- B_{\vec{h}}(y) = 0 \ ,
\elea(Bvh)
where the functions $A_{\vec{h}}, B_{\vec{h}}, 
C_{\vec{h}}, D_{\vec{h}}$ of $y$ are the following
matrix elements,  
$$
\left( \begin{array}{cc}
-A_{\vec{h}}(y)  &  B_{\vec{h}}(y)  \\
C_{\vec{h}}(y)  & - D_{\vec{h}}(y)    
\end{array} \right) :
= \prod_{j=0}^{L-1}  \left( \begin{array}{cc}
- a^N_j  &  y b^N_j  \\
 y c^N_j  &-d^N_j    
\end{array} \right) \ .
$$
Note that  
${\cal B}_{\vec{h}} $ is a double cover of 
$y$-line, and  
${\cal C}_{\vec{h}} $ is a
$(\ZZ_N)^{L+1}$-branched cover of 
${\cal B}_{\vec{h}}$. The automorphisms $\tau_\pm$ 
generate a  covering
transformation group of ${\cal C}_{\vec{h}} $ over 
${\cal B}_{\vec{h}}$. In this section, we shall
only consider the case 
$$
L = 3 \ , \ \ a_0=d_0=0 \ , \ b_0=c_0 = 1 \ ,
$$
and we assume the variables  $h_1, h_2$ to be
generic. The expression of $T(x)$ is given by
$$
 T( x)  = x^2 (c_1a_2 X\otimes Z \otimes Y + 
 a_1b_2 Z \otimes Y
\otimes X  + b_1d_2 Z\otimes X \otimes I + 
 d_1c_2 X \otimes I
\otimes Z ) ,
$$
equivalently, $x^{-2} D^{\frac{-1}{2}}  T( x)$ is
equal to the  Hofstadter Hamiltonian (\req(HofH)) with
$U, V$ given by (\ref{UV}) and 
 $\mu, \nu,
\alpha,
\beta$ related to 
$h_1, h_2$  by 
$$
\mu^2 = qb_1c_1a_2d_2 \ , \ \alpha^2 =
q^{-1}b_1c_1^{-1}a_2^{-1}d_2 \ , \ \nu^2 = q
a_1d_1b_2c_2 \ , \ 
\beta^2= q^{-1}a_1^{-1}d_1b_2^{-1} c_2 \ .
$$
The curve ${\cal B}_{\vec{h}}$ is defined by (\req(Bvh))
with 
$$
\begin{array}{ll}
A_{\vec{h}}(y)= -y^2( c^N_1a_2^N+ d_1^N c_2^N) \
, & 
B_{\vec{h}}(y)=y(y^2c_1^Nb_2^N+d_1^Nd^N_2 ) \ ,
\\  C_{\vec{h}}(y)=y(y^2b_1^Nc_2^N + a^N_1a_2^N )
\ ,& 
D_{\vec{h}}(y)=-y^2(a_1^N b^N_2+ b_1^Nd_2^N ) \ .
\end{array}
$$
By factoring out the $y$-component, we consider
only the main irreducible component of ${\cal
B}_{\vec{h}}$, denoted by
\be
{\cal B}: \ (y^2b_1^Nc_2^N + a^N_1a_2^N ) 
\eta^2 + ( a_1^N b^N_2+ b_1^Nd_2^N 
- c^N_1a_2^N- d_1^N
c_2^N) y \eta  -
(y^2c_1^Nb_2^N+d_1^Nd^N_2 ) = 0 \ ,
\ele(B)
which is a double-cover of $y$-line with
four branched points, hence it defines an elliptic
curve.  For the curve ${\cal
C}_{\vec{h}}$, 
the variables $\xi_0$ and $ \xi_1$ are related by
$\xi_0^N =
\xi_1^{-N}$. This implies that ${\cal
C}_{\vec{h}}$ consists of
$N$ irreducible components,  each one is
isomorphic to the same curve ${\cal W}$
defined by the equations in the variable
$p=(x,\xi_0, \xi_2)$, 
$$
{\cal W} : \ \ \xi_0^{-N} = \frac{-\xi_2^Na_1^N+
x^Nb_1^N}{x^N\xi_2^Nc_1^N-d_1^N} \ , \ \ \
\xi_2^N= \frac{-\xi_0^Na_2^N+
x^Nb_2^N}{x^N\xi_0^Nc_2^N-d_2^N} \ .
$$
It is easy to see that ${\cal W}$ is a $N^3$-fold
(branched) cover of the elliptic curve (\req(B)), 
and the genus of ${\cal W}$ is 
$6N^3-6N^2+1$.  We shall label the irreducible components of
${\cal C}_{\vec{h}}$ by $s \in
\ZZ_N$ and denote the elements of ${\cal
C}_{\vec{h}}$ by $(p, s)$ with
$p
\in {\cal W}$, whose 
$\xi_1$-coordinate is given by $\xi_0\xi_1=
q^{2s-1}$. The relation (\req(T|p)) now becomes
$$
T(x)|p, s \rangle = |\tau_-(p), s-1 \rangle
q^{2s-1}\Delta_-(p) +  |\tau_+(p), s-1\rangle q^{2s}
\Delta_+(p) \ , 
$$
where $\tau_{\pm}$  are transformations of
of ${\cal W}$ defined by 
(\ref{DelTau}), with the coordinates only
involving 
$(x, \xi_0, \xi_2)$, and $\Delta_\pm$ are the
following functions on
${\cal W}$,
\begin{eqnarray*}
\Delta_-(p )  = &
 - x
\xi_0^{-1} ( x
\xi_2 c_1 -d_1 ) ( x
\xi_0 c_2 - d_2 )   \ , & \\
\Delta_+(p )  = & x \xi_2
\frac{
(a_1d_1-x^2b_1c_1)(a_2d_2-x^2b_2c_2)}{(\xi_2a_1
-xb_1)(\xi_0a_2-xb_2)} \ , & p=(x, \xi_0, \xi_2) \in
{\cal W} \ . 
\end{eqnarray*}
By averaging the vectors $|p, s \rangle$ over an element
$p$ of ${\cal W}$, one defines the following Baxter
vector on 
${\cal W}$
$$
|p \rangle:= \frac{1}{N} \sum_{s=0}^{N-1} |p, s \rangle
q^{s^2} 
\ .
$$
The action of the transform matrix $T(x)$ on $|p,
s\rangle$ can be descended to the Baxter vector of
${\cal W}$ as follows:
\bea(l) 
x^{-2}T(x)|p\rangle = |\tau_-(p)\rangle
\widetilde{\Delta}_-(p) +  |\tau_+(p)\rangle 
\widetilde{\Delta}_+(p) \ ,
\elea(THof)
where $\widetilde{\Delta}_{\pm}$ are the
functions of
${\cal W}$,
\begin{eqnarray*}
\widetilde{\Delta}_-(x, \xi_0, \xi_2 ) =&
 \frac{( x
\xi_2 c_1 -d_1 ) ( x
\xi_0 c_2 - d_2 )}{- x \xi_0}, \\
\widetilde{\Delta}_+(x, \xi_0, \xi_2 ) =& 
\frac{ \xi_2
(a_1d_1-x^2b_1c_1)(a_2d_2-x^2b_2c_2)}{x (\xi_2a_1
-xb_1)(\xi_0a_2-xb_2)} . 
\end{eqnarray*}
By the following component expression  of the Baxter
vector on
${\cal C}_{\vec{h}}$,
$$
\langle k_0, k_1, k_2 |p, s \rangle q^{s^2} = 
q^{(s-k_0-k_1)^2} q^{  -2
k_0^2- 2k_0k_1 -k_1^2  +k_1 }
\prod_{i=1}^{k_1} \frac{\xi_0(-\xi_2a_1\omega^i + xb_1)}
{\xi_2xc_1\omega^i-d_1}
\prod_{j=1}^{k_2}
\frac{-\xi_0a_2\omega^j+xb_2}
{\xi_2(\xi_0xc_2\omega^j-d_2)},
$$
the Baxter vector on ${\cal W}$ 
is given by 
\begin{eqnarray}
\langle k_0, k_1, k_2 |p \rangle= \frac{\sum_{n=0}^{N-1}
q^{n^2}}{N} q^{  -2
k_0^2- 2k_0k_1 -k_1^2  +k_1 }
\prod_{i=1}^{k_1} \frac{\xi_0(-\xi_2a_1\omega^i + xb_1)}
{\xi_2xc_1\omega^i-d_1}
\prod_{j=1}^{k_2}
\frac{-\xi_0a_2\omega^j+xb_2}{\xi_2(\xi_0xc_2\omega^j-d_2)}
. \label{compp}
\end{eqnarray}
Note that the $k_j$s in the above formula are integers
modular $N$. Each product-term of the
right hand side means the one for  
a positive integer $k_j$ representing its 
class in $\ZZ_N$.  For an eigenvector $\langle \varphi|$
in $\stackrel{3}{\otimes} \CZ^{N*}$ of the operator
$x^{-2}T(x)$
 with the eigenvalue 
$\lambda$, by (\req(THof)), the function 
$
Q(p) : = \langle  \varphi |
p \rangle  $ 
 of ${\cal W}$ satisfies the following Bethe equation,
\bea(l)
\lambda Q(p) = Q(\tau_-(p)) \widetilde{\Delta}_-(p) + 
Q(\tau_+(p)) \widetilde{ \Delta}_+(p) \ , \ \ \lambda
\in \CZ \ .
\elea(BEHof)
The above equation possesses a $\ZZ_2$-symmetry  with
respect to the following  involution of ${\cal W}$,
$$
\sigma: {\cal W} \longrightarrow {\cal W} \ , \ \ 
p= (x, \xi_0, \xi_2) \mapsto  \sigma(p) = (-x, -\xi_0,
-\xi_2)
\ .
$$
In fact, the commutativity of $\sigma$ and 
$\tau_\pm$, and the $\sigma$-invariant property of  
$\widetilde{\Delta}_\pm(p )$ are easily seen. Then
by (\ref{compp}),  Baxter vector $|p\rangle$ is
invariant under $\sigma$, i.e.,
$|p\rangle = |\sigma(p)\rangle $ for $p \in {\cal W}$,
which implies that  $Q(p)$ is a
$\sigma$-invariant function. Furthermore the
rational function $Q(p)$ has  the poles
contained in the following divisor, 
$$
\xi^{N-1} \prod_{i=1}^{N-1}(\xi_2 x c_1 \omega^i-d_1)
(\xi_0 x c_2 \omega^i-d_2) = 0 \ ,
$$
in particular, it is regular at
$x=0, \infty$. Note that the finite values of $Q(p)$ at
$x=0,
\infty$  are consistent
with  the asymptotic values of $\widetilde{\Delta}_\pm$
at $x=0, \infty$ in the equation (\req(BEHof)),
\begin{eqnarray*}
 \widetilde{\Delta}_\pm (x , \xi_0, \xi_2 )  = &
 \pm x^{-1}
\xi_0^{-1} d_1  d_2 + O (1)    &{\rm as } \ \ x
\rightarrow 0 \ , 
 \\
\widetilde{\Delta}_\pm ( x , \xi_0, \xi_2 )  = &
 \pm x  \xi_2 c_1 c_2  + O (1)   & {\rm as } \ \ x
\rightarrow \infty \ .
\end{eqnarray*}
With the $x$-coordinate,
${\cal W}$ is a $2N^2$-cover over the
$x$-line,  unramified at points over $x=0,
\infty$ whose $(x, \xi_0, \xi_2)$-coordinates
are given by
\begin{eqnarray}
0_{i,i'}^\pm =\pm  (0,  q^i
\sqrt{\frac{d_1d_2}{a_1a_2}}, 
q^{i'}\sqrt{\frac{d_1a_2}{a_1d_2}} ) \ ,
\ \ 
\infty_{i,i'}^\pm = \pm 
(\infty, q^i\sqrt{\frac{c_1c_2}{b_1b_2}} ,  
q^{i'} \sqrt{\frac{c_1b_2}{b_1c_2}} ) \ , \ \ \ 
i, i' \in \ZZ_N \ . \label{0infty}
\end{eqnarray}
Consider the
$D$-eigenspace decomposition
of 
$\stackrel{3}{\otimes} \CZ^{N*} = \bigoplus_{l\in
\ZZ_N} \EZ_3^l$ in Sect. 6.
The evaluation on the Baxter vector of ${\cal W}$ gives
rise  to the following linear transformation,
$$
\varepsilon_l : \EZ_3^l \longrightarrow \{ {\rm 
rational \ functions \ of \ } {\cal W} \} , \ v 
\mapsto \varepsilon_l(v)(p) := \langle v|p\rangle \ ,
\ \  {\rm for} \ l \in \ZZ_N \ .
$$ 
\begin{theorem}\label{thm:Hof}
For $l \in \ZZ_N$, the
linear map $\varepsilon_l$ is injective, hence it induces
an identification of $\EZ_3^l$ with a
$N^2$-dimensional functional space of
${\cal W}$.  
\end{theorem}
{\bf Proof.} Define the following vectors in 
$\stackrel{3}{\otimes} \CZ^{N*}$,
\begin{eqnarray*}
\langle \psi_{k_0, k_1, k_2}|= & (\sum_{n=0}^{N-1}
q^{n^2})^{-1} \sum_{k' \in \ZZ_N}
\omega^{k_1 k'}\langle k_0, k', k_2 | \ , \\
\langle \phi_{j_0, j_1, j_2} | = & \sum_{k \in \ZZ_N} 
q^{2k(-j_0+j_1 + j_2)- k(k-1)} \langle  \psi_{j_1-k, k,
j_2-k} | \ , 
\end{eqnarray*}
where $k_i, j_i \in \ZZ_N$. We have 
$$
\begin{array}{l}
\langle  \psi_{k_0, k_1, k_2}| Z \otimes X \otimes I =
\omega^{k_0 + k_1} \langle \psi_{k_0, k_1, k_2}| \ , \\ 
\langle  \psi_{k_0, k_1, k_2}| X \otimes I \otimes Z =
\omega^{k_2} \langle  \psi_{k_0-1, k_1, k_2} | \ , \\
\langle  \psi_{k_0, k_1, k_2}| I \otimes Z \otimes X =
 \langle  \psi_{k_0, k_1+1, k_2-1} | \ .
\end{array}
$$
By (\req(3D)), one has 
\begin{eqnarray*}
\langle  \psi_{k_0, k_1, k_2}|D  = q^{-1+2(k_0 + k_1+
k_2)} \langle 
\psi_{k_0-1, k_1+1, k_2-1} | \ , \ \ \
\langle \phi_{j_0, j_1, j_2}| D = q^{-1+2 j_0}\langle 
\phi_{j_0, j_1, j_2}| \ .
\end{eqnarray*}
For a given $l \in\ZZ_N$, let $j_0$ be the element in
$\ZZ_N$ defined by $q^l= q^{-1+2j_0}$. Then  the vectors
$\langle
\phi_{j_0, j_1, j_2}|$ with
$  j_1, j_2 \in \ZZ_N$  form a basis of $\EZ_3^l$. By
(\ref{compp}), we have 
\begin{eqnarray*}
\langle \psi_{k_0, k_1, k_2}|p\rangle = N^{-1} q^{  -2
k_0^2 }
\sum_{k' \in \ZZ_N} 
 q^{ -k'^2+ (2k_1 - 2k_0   +1)k' }
\prod_{i=1}^{k'} \frac{\xi_0(-\xi_2a_1\omega^i + xb_1)}
{\xi_2xc_1\omega^i-d_1}
\prod_{j=1}^{k_2}
\frac{-\xi_0a_2\omega^j+xb_2}
{\xi_2(\xi_0xc_2\omega^j-d_2)} \ ,
\end{eqnarray*}
hence
$$
\begin{array}{ll}
\langle \phi_{j_0, j_1, j_2}|p\rangle =&  N^{-1}  q^{ 
-2j_1^2 }
 \sum_{k,k' \in \ZZ_N} 
q^{-3k^2+ k(-2j_0+6j_1 + 2j_2+1) -k'^2+4k'k+ (- 2j_1  
+1) k' } \\
& \prod_{i=1}^{k'}
\frac{\xi_0(-\xi_2a_1\omega^i + xb_1)}
{\xi_2xc_1\omega^i-d_1}
\prod_{j=1}^{j_2-k}
\frac{-\xi_0a_2\omega^j+xb_2}
{\xi_2(\xi_0xc_2\omega^j-d_2)} \ .
\end{array}
$$
Set $p = 0_{i,i'}^\pm$ defined in
(\ref{0infty}). By the relations of their
$(\xi_0,
\xi_2)$-coordinates, $
\xi_0\xi_2a_1 d_1^{-1} = q^{i+i'}$ ,  
$\xi_0\xi_2^{-1}a_2d_2^{-1} = q^{i-i'}$, 
we obtain
\begin{eqnarray*}
\langle \phi_{j_0, j_1, j_2}|0_{i,i'}^\pm\rangle &=&
N^{-1} q^{  -2j_1^2 +j_2^2+j_2(1+i-i')}
 \sum_{k, k' \in \ZZ_N} 
q^{-2k^2 + k(-2j_0+6j_1 -i+i')  }
 q^{   k'( -2j_1 +2+i+i'+4k) } \\
&= & q^{j_0(1- j_1) - 2 j_1 + \frac{j_1^2}{2}  +j_2
+j_2^2    +
\frac{-j_1( 3i + i')+2j_2(i-i') }{2}+   \frac{ j_0(i+
i')}{2} -\frac{( i+i'+2)( - i+ 3i'+2)}{8} }
\end{eqnarray*}
hence
\begin{eqnarray}
q^{j_0(-1+ j_1 )+ 2 j_1 - \frac{j_1^2}{2} -j_2
-j_2^2 }
\langle
\phi_{j_0, j_1, j_2}|0_{i,i'}^\pm\rangle q^{\frac{(
i+i'+2)( - i+ 3i'+2)-4j_0( i+ i')}{8} } = q^{
 \frac{ i(-3j_1+2j_2)
-i'(j_1+2j_2)}{2}
 } \ . \label{phi0}
\end{eqnarray}
As the correspondence, $(j_1, j_2) \mapsto (-3j_1+2j_2,
-j_1-2j_2)$, defines an automorphism of $\ZZ_N^2$, the
relation (\ref{phi0}) gives rise to an isomorphism
between
$\EZ_l$ and the space of Baxter vectors 
$0_{i,i'}^+$s (or equivalently $0_{i,i'}^-$s). This 
implies the injectivity of $\varepsilon_l$.
$\Box$ \par \vspace{0.2in} \noindent
By the discussion in Sect. 6,  $\EZ_3^l$ is equivalent to $N$ copies of 
the standard representation as the
Heisenberg algebra ${\cal O}_3$-modules. Hence by
Theorem \ref{thm:Hof}, there exists an ${\cal
O}_3$-module structure
on $\varepsilon_l(\EZ_3^l)$, inherited from the 
representation space $\EZ_3^l$. The mathematical
structure of the functional
space $\varepsilon_l(\EZ_3^l)$ by incorporating
the divisor theory of Riemann surfaces into the
Heisenberg algebra representation remains an
algebraic geometry problem for further study.

\section{ Conclusions and Perspectives }
We follow the framework in \cite{FK} by
the quantum integrable method to study
the diagonalization problem of some
Hofstadter-like models. Through the Baxter vector
of the spectral curve, the study of diagonalizing
a Hamiltonian with the rational magnetic flux is
reduced to the problem of a certain
"Strum-Liouville-like" difference equation on the
curve, called the Bethe equation of the
associated model. The spectral curve is in
general with a large genus, and the relations
among zeros and poles for a solution of the Bethe
equation yields a system of algebraic equations.
Such systems of relations among zeros of
the Bethe polynomial solution on a rational
spectral curve are usually referred to the  Bethe
ansatz equation in literature. For certain
 models of physical interest, e.g.,  the
Hofstadter Hamiltonian  (\req(HofH)), the study
of the high genus spectral curve is a necessary
step in solving the spectrum problem through the
algebraic Bethe-ansatz-technique. In this paper,
we clarify some finer mathematical manipulations
in
\cite{FK}, then  go through all the delicate
points one must consider in order to
obtain the explicit Bethe solutions. A careful
analysis of the mathematical nature of the
the Bethe equation reveals the vital role of
algebraic geometry in a
thorough understanding of the Bethe ansatz
method, so is the same with the need to obtain
the physical answer of the associated model. For
this reason, we have examined in this paper the
Bethe ansatz equation in the context of algebraic
geometry, even in
the degenerated  rational spectral
curve case in order for gaining the mathematical
insight of the Bethe equation. We further
extend the approach to some more general
situations.  Above all, we have endeavoured to
present a clear and  self-contained account of
the theory, and hope to have elucidated the
mathematical structure of the Bethe-ansatz-style
method in the physical literature. The main
content of this paper is in the discussions
after Sect. 5, where the detailed mathematical
derivation and analysis comparable to physical
considerations are presented. The topics
between Sect. 4 and Sect. 7 are devoted to the
degenerated case, where the Bethe equations are
related to models with the rational  spectral
curve. With an explicit gauge choice, we have
conducted the mathematical investigation of the
Bethe equation and obtained the complete Bethe
solutions for all sectors in Sect. 5. With these
mathematical results, we are able to further
advance the study of the relevant physical
problems, namely, the "degeneracy" of eigenstates
of the transfer matrix to the Bethe solutions and
the thermodynamic limit discussion in Sects. 6 and
7. Furthermore in Sect. 6, the
explicit calculation we have performed for the
Bethe solutions, when specializing on one
particular sector, provides the results parallel
to those using the usual Bethe ansatz method in
\cite{FK}. Meanwhile, the finding of some
non-physical Bethe ansatz solutions in other sectors
supports the  justification of our approach of the
problem.  The method we employ here can apply equally
well to any number of size
$L$. However, to keep things simple, we restrict
our attention in this
paper  only to the case $L
\leq 3$ in the discussion of Bethe solutions; the
analysis on
$L=1,2$ are mainly for the mathematical purpose to pave
the way of discussing the models with a higher 
size $L$. The Hofstadter-like model is related to
$L=3$, of which the Bethe ansatz equation has
been  mathematically discussed in details here. 
For $L=4$, it is expected that the problem should
be closely related to the discrete quantum
pendulum  by the works of
\cite{BKP, KKS}. The results  we have obtained in this
paper through the  Bethe equation
approach strongly indicate  a promising direction to the
spectra problem of other models, e.g., the
discrete quantum pendulum.  For the thermodynamic limit
discussion,  the analysis in Sect. 7 on the special
Hofstadter-like Hamiltonian (\req(GSHof)) for a
generic
$q$ shows that  the Bethe relation on the
spectrum and eigenstates we proposed are in
accordance with the diophantine approximation
process of an irrational flux. The comparison of
our Bethe ansatz method with the
$C^*$-algebra approach of semiclassical analysis
employed in \cite{B} for the multifractal spectrum
structure is a fascinating problem. We plan to address
the question of a such program elsewhere.

For the original Hofstadter model (\req(HofH)),  the
Bethe equation is formulated as a "difference"
equation of functions on a high genus spectral
curve. In Sect. 8, we have made a primary investigation
on its solutions . As the spectral curve is a Galois
abelian cover of an elliptic curve with the covering
group determined by the order 
$N$ of the rational flux, it would be essential to have a
detailed algebraic geometry study of such a high genus
curve in cooperation with the Bethe  solutions,
so that the base elliptic curve and the classical
elliptic functions could be engaged in the
theory.  With our finding in the rational
spectral curve case as a guidance for some
appropriate direction of calculation, the
analysis we  have made in this paper will serve
as a basis for further study of the Hofstadter
model through  the elliptic curve techniques in
algebraic geometry.  This approach would allow us
to follow a similar path as in the
rational spectral curve case for the study of the
spectrum problem of the Hofstadter model.  This
rich structure requires further study, and  a
scheme along this line of interpretation is  under
current investigation. Indeed, we hope that
our efforts would eventually shed  new light on
the role of algebraic geometry in  exactly
solvable integrable models. In this paper we
restrict our attention only to certain
Hofstadter-like models, and  leave possible
generalizations and applications to future work.

\section*{Acknowledgements}
One of the authors, S.S. Roan, would like to thank 
R. Seiler and J.  Kellendonk for the stimulating
conversations on the Hofstadter model during their
encounters in U.C. Berkeley, U.S.A., in the spring of
1998. He would acknowledge informative discussions and
communications with B. M. McCoy on R. Baxter's work. We
would also like to thank M. Jimbo for the inspiring
discussions during the final stage of this paper. This
work was reported by the second author in the workshop
"Quantum Algebra and Integrabilty" CRM Montreal, Canada,
April 2-15 2000, and a part of subject of an Invited
Lecture at APCTP-Nankai Joint Symposium on "Lattice
Statistics and Mathematical Physics", Tianjin, China,
Octorber 7- 11, 2001, to which he would like to thank
for their invitation and hospitality.

\end{document}